\documentclass[showpacs,preprintnumbers,
superscriptaddress,amsmath]{revtex4}

\usepackage{epsfig}

\newcommand{\PRD}[3]{Phys.\ Rev.\ D\ {\bf #1},\ #2 (#3)}

\newcommand{\be}{\begin{equation}}
\newcommand{\ee}{\end{equation}}
\newcommand{\bea}{\begin{eqnarray}}
\newcommand{\eea}{\end{eqnarray}}
\newcommand{\ba}[1]{\begin{array}{#1}}
\newcommand{\ea}{\end{array}}

\newcommand{\eq}[1]{Eq.\,(\ref{#1})}

\newcommand{\eqq}[2]{Eqs.\,(\ref{#1}) and (\ref{#2})}

\newcommand{\eeq}[1]{(\ref{#1})}

\begin{document}

\title{Explicit form of quark propagators and gluon self-energies in neutral three-flavor color superconductor}

\author{H.\ Malekzadeh}
\affiliation{Institut f\"ur Theoretische Physik,
Westf\"alische Wilhelms-Universit\"at M\"unster, Wilhelm-Klemm-Str.\ 9
48149 M\"unster, Germany\\
E-mail: malekzadeh@uni-muenster.de}

\date{\today}

\begin{abstract}
Enforcing color and electric charge neutrality conditions on the three-flavor color superconducting matter, I derive the explicit form of the quark propagators and the gluon self-energies for the gapless and the ordinary color-flavor-locked phases.
\end{abstract}

\maketitle

\section{Introduction}

Single-gluon exchange between two quarks is attractive in the color-antitriplet channel. Therefore, sufficiently cold and dense quark matter is a color superconductor (CSC) \cite{bailinlove}. Since at one-loop order the gluon loop is $\sim g^2T^2$ whereas the quark loop is $\sim g^2\mu^2$, the effects of the gluon loops are very negligible and the main contribution results from the quark loops \cite{son, meissner2, 2frischke, 3cf}.

There are many different color-superconducting phases. For the quark chemical potentials much larger than the strange quark mass, $m_s$, the ground state of the color-superconducting quark matter is the so-called color-flavor-locked (CFL) phase \cite{arw}. At smaller densities, however, the so-called 2SC phase is the dominant phase.

The Meissner mass is computed for the 2SC phase in Refs.\,\cite{meissner2, 2frischke}, and that for the CFL phase in Refs.\,\cite{sonstephanov, 2sc}. Furthermore, the full energy-momentum dependence of the one-loop gluon self-energy for the 2SC phase is investigated in Ref.\,\cite{dirkigor}. The respective calculations for the CFL phase are done in Ref.\,\cite{3cf}. It is shown that square of the Meissner mass in the 2SC phase is negative, hence, this phase suffers from the so-called chromomagnetic instability \cite{chromo2sc}. When the neutrality condition, which is a crucial criterion for the compact stars \cite{neutral1, neutral2}, is enforced, the scenario changes drastically. Under this condition, the gapless phases are produced \cite{gapless}. However, it is realized that the gapless 2SC (g2SC) phase as well as the gapless CFL (gCFL) phase are unstable. This instability is analogous to that for the 2SC phase \cite{chromog2sc, chromocfl}. There are studies on the other phases as well. However, those phases are either unstable or energetically disfavored or could be the ground state only at a very tiny region of the phase diagram, e.g. the ${\rm CFL-K}^0$ phase \cite{kcfl}, the single-flavor phase \cite{singleflavor}, the secondary-pairing phase \cite{secondarypairing}, the mixed phase \cite{mixedphases}, the LOFF phase \cite{loffff}, the gluonic phase \cite{gluonic, gluonic1, chromofurther}, and the inert phases \cite{inertphases}.

It is known that the matter in the CFL phase as well as the 2SC phase contains a new kind of the collective modes, the so-called ``light'' plasmon modes \cite{plasmon, casalbuoni, 3cf}. In contrast to the ordinary plasmon, there is no trace of the light plasmon modes in the normal phase. The plasma frequency $\omega_p$ is proportional to the value of the chemical potential and is independent of temperature and/or the color superconducting gap. However, the light plasmon has very different properties. Although it is a massive excitation too (that is why we also call it a plasmon), its mass $m_\Delta$ is of the color superconducting gap order, more precisely, $1.362|\Delta_T|< m_\Delta < 2|\Delta_T|$.

In the long wavelength limit $|\bar{q}|\rightarrow 0$, the light plasmons are stable with respect to the decays into the quark type quasiparticles. Besides, the energy of this mode is a monotonically increasing function of momentum $|\bar{q}|$ \cite{plasmon}. Therefore, there is a critical value for the wavelength, $1/|\Delta_T|$, at which the energy of the light plasmon becomes equal to the threshold of the quasiparticle pair production. The excitations with the wavelengths shorter than the critical value (and the energy larger than $2|\Delta_T|$) can decay into quasiparticles and, therefore, must have a relatively large width. It means that these new type of plasmon modes have two characteristic plasma frequencies, $m_\Delta$ and $2|\Delta_T|$. The stable (narrow width) modes live only in $m_\Delta <q_0< 2 |\Delta_T|$ window of the energies.

The manifestation of the chromomagnetic instability in the two-flavor CSC is quite different for the 2SC and the g2SC phases. In the 4-7th channels of the 2SC phase as well as the g2SC phase, the chromomagnetic instability reflects the typical Bose-Einstein condensation phenomena: while at subcritical values of $\delta\mu$, there are the light plasmons with the gap squared at $0 < \mathcal{M}^2 \leq \Delta^2 \ll \mu^2$, at subcritical values $\delta\mu>\Delta/\sqrt{2}$, the plasmons become tachyonic with $\mathcal{M}^2<0$. The spectrum of the plasmons in the 8th channel is quite different. There are no light plasmons at all in the 2SC phase in that channel. On the other hand, in the g2SC phase, there is a gapless tachyonic plasmon \cite{gluonic1}.

The gapless tachyonic instability may appear in the three-flavor CSC too. The instability in this case is generated in the channels with the quantum numbers of $A^{(1)}, A^{(2)}$ gluons and a linear combination of diagonal $A^{(3)}, A^{(8)}$ gluons and photon $A^{(\gamma)}$, cf. Ref.\,\cite{gluonic1}. In the gCFL phase, the instability must be similar to the chromomagnetic $A^{(8)}$ instability in the g2SC phase \cite{gluonic3}.

The chromomagnetic instability of the gCFL phase can be resolved if the light plasmon modes condensate. In order to investigate this possibility one has to know the gluons polarization tensor in the neutral three-flavor CSC phase. These calculations need the neutral quark propagators. Whether the light plasmon modes condensate or not the scenario of the instability in the gCFL phase undergoes some change. In addition, the neutral quark propagators and gluon self-energies are needed to calculate many phenomena related to the compact stars, e.g. the effects of the magnetic field on the gluon spectrum, the thermodynamic quantities and etc. Once these calculations are performed, the main complication of the analytical investigations in the neutral three-flavor CSC eases. Here, I focus on the propagators and the self-energies; the light plasmon condensate will be studied elsewhere \cite{future}.

This paper is organized as follows. In the next section, Sec.\,\ref{sec1}, including the neutrality conditions, I calculate the quark propagators in the three-flavor CSC. One can find the form of the propagators in the CFL phase in Sec.\,\ref{sec2}. In section \ref{neutralcfl}, using the quark propagators, I derive the gluon polarization tensors. Afterwards, in Sec.\,\ref{sec5}, I restrict the results to the CFL phase and find the gluon self-energies in this phase too.

In this paper I use the natural units, $\hbar=c=k_B=1$, and work in Euclidean space-time ${\bf R}^4\equiv V/T$ where $V$ is the volume and $T$ is the temperature of the system. I use the Minkowski notation for 4-vectors, with the metric $g^{\mu\nu}={\rm diag}(+,-,-,-)$.

\section{Quark propagators}\label{sec1}

In the color superconducting quark matter, the ground state is a condensate of the quark Cooper pairs. In the mean-field approximation the inverse quark propagator matrix is
\be\label{propinverse}
S^{-1}(K)=\left(
\begin{array}{cc}
(G_0^+)^{-1}(K) & \phi^- \\
\phi^+ & (G_0^-)^{-1}(K)
\end{array}\right)\;,
\ee
where $G_0^\pm$ is the propagator for noninteracting quarks (upper sign) or charge conjugate quarks (lower sign) with the quark four-momentum $K\equiv(k_0, k)$ and $\phi^\pm$ is the gap matrix \cite{pisarischk1}. The propagator $G_0^\pm$ has the following form
\bea\label{qprop}
(G_0^+)^{-1}(K)=(\gamma^\mu K_\mu \pm \mu\gamma^0-m)\,,
\eea
where $\mu$ is the quark chemical potential, $m$ is the quark mass matrix, and $\gamma^\mu$ is the Dirac matrix. The gap matrices are defined through
\bea
\phi^+\sim\, <\psi_C\,\bar{\psi}>\,\,\,,\,\,\, \phi^-\sim\, <\bar{\psi}_C\,\psi>
\eea
with the quark field
\bea
\psi=(\,\psi^u_r\,,\, \psi^d_r\,,\, \psi^s_r\,,\, \psi^u_g\,,\, \psi^d_g\,,\, \psi^s_g\,,\, \psi^u_b\,,\, \psi^d_b\,,\, \psi^s_b\,)^T\;.
\eea
The lower indices ($r, g,$ and $b$) stand for the quark color and the upper ones ($u$, $d$, and $s$) for the quark flavor. The charge conjugate quark field is defined via
$\psi_C=C\bar{\psi}^T$, where $C=i\gamma^2\gamma_0$. The color-flavor structure of the quark propagator in Eq.\,(\ref{qprop}) is given by
\bea\label{cfprop}
\left[G_0^\pm(K)^{-1}\right]^{fg}_{ij}=\gamma^\mu K_\mu \delta^{fg}\delta_{ij} + \gamma_0\,\mu^{fg}_{ij}\;.
\eea
where, for simplification, we pulled the quark mass matrix $m$ into the quark chemical potential $\mu^{fg}_{ij}$. This does not cause any changes in the result. In order to make the system color charge neutral one has to make the tadpoles of the system vanish. In the NIL model this is done by introducing the relevant chemical potentials. This, in the three-flavor CSC, leads to adding $\mu_3$ and $\mu_8$ to the quark chemical potential \cite{neutral3}. The system has to be electric charge neutral as well. For the latter the quark electrical chemical potential $\mu_Q$ related to the quark electric charge
\bea
Q_F=\left(
\begin{array}{ccc}
\frac{2}{3} & 0 & 0 \\
0 & -\frac{1}{3} & 0 \\
0 & 0 & -\frac{1}{3}
\end{array}\right)
\eea
must be taken into account. In addition, considering the large value of the strange quark mass, we can set the up and the down quark masses to zero. In the end, the quark chemical potential finds the following form
\bea
\mu^{fg}_{ij}= \mu\,\delta^{fg}\delta_{ij} + \mu_Q\, Q_F^{fg}\delta_{ij} + \mu_8\, (T_8)_{ij}\delta^{fg} + \mu_3\, (T_3)_{ij}\delta^{fg} - \frac{m_s^2}{2\mu}\delta^{f3}\delta^{g3}\delta_{ij}\;,
\eea
which is a $9\times 9$ matrix in the color $[ij]$ and the flavor $[fg]$ spaces. Here, $T_3$ and $T_8$ are the Gell-Mann matrices. The nonzero components of $\mu^{fg}_{ij}$ are the diagonal entries
\bea\label{chemical}
\mu^{11}_{11}&\equiv&\mu_{ru} = \mu - \frac{2}{3}\mu_e + \frac{1}{2}\mu_3 + \frac{1}{3}\mu_8\;,\nonumber\\
\mu^{22}_{22}&\equiv&\mu_{gd} = \mu + \frac{1}{3}\mu_e - \frac{1}{2}\mu_3 + \frac{1}{3}\mu_8\;,\nonumber\\
\mu^{33}_{33}&\equiv&\mu_{bs} = \mu + \frac{1}{3}\mu_e - \frac{2}{3}\mu_8 - \frac{m_s^2}{2\mu}\;,\nonumber\\
\nonumber\\
\mu^{11}_{22}&\equiv&\mu_{gu} = \mu - \frac{2}{3}\mu_e - \frac{1}{2}\mu_3 + \frac{1}{3}\mu_8\;,\nonumber\\
\mu^{22}_{11}&\equiv&\mu_{rd} = \mu + \frac{1}{3}\mu_e + \frac{1}{2}\mu_3 + \frac{1}{3}\mu_8\;,\nonumber\\
\nonumber\\
\mu^{33}_{11}&\equiv&\mu_{rs} = \mu + \frac{1}{3}\mu_e + \frac{1}{2}\mu_3 + \frac{1}{3}\mu_8 - \frac{m_s^2}{2\mu}\;,\nonumber\\
\mu^{11}_{33}&\equiv&\mu_{bu} = \mu - \frac{2}{3}\mu_e - \frac{2}{3}\mu_8\;,\nonumber\\
\nonumber\\
\mu^{22}_{33}&\equiv&\mu_{bd} = \mu + \frac{1}{3}\mu_e - \frac{2}{3}\mu_8\;,\nonumber\\
\mu^{33}_{22}&\equiv&\mu_{gs} = \mu + \frac{1}{3}\mu_e - \frac{1}{2}\mu_3 + \frac{1}{3}\mu_8 - \frac{m_s^2}{2\mu}\;,
\eea
Inserting the chemical potentials into the inverse of the quark propagator, Eq.\,(\ref{cfprop}), the nonzero components of the inverse bare propagators in color and flavor spaces are
\bea
[(G_0^\pm)^{-1}]^{11}_{11}\,,\,[(G_0^\pm)^{-1}]^{11}_{22}\,,\,[(G_0^\pm)^{-1}]^{11}_{33}\,,\nonumber
\eea
\bea
\,[(G_0^\pm)^{-1}]^{22}_{11}\,,\,[(G_0^\pm)^{-1}]^{22}_{22}\,,\,[(G_0^\pm)^{-1}]^{22}_{33}\,,\nonumber
\eea
\bea
\,[(G_0^\pm)^{-1}]^{33}_{11}\,,\,[(G_0^\pm)^{-1}]^{33}_{22}\,,\,[(G_0^\pm)^{-1}]^{33}_{33}\,.
\eea
These components are just entries of the matrix composed of the color and the flavor spaces. Therefore, one can find their inverse, which is in Dirac space, without having to invert the color-flavor-matrix. Hence the only nonzero components of the bare propagator are
\bea\label{comprop}
(G_0^\pm)^{11}_{11}\,,\, ((G_0^\pm)^{11}_{22}\,,\, (G_0^\pm)^{11}_{33}\,,\, (G_0^\pm)^{22}_{11}
\,,\, G_0^\pm)^{22}_{22}\,,\, (G_0^\pm)^{22}_{33}\,,\, (G_0^\pm)^{33}_{11}\,,\, (G_0^\pm)^{33}_{22}\,,\, (G_0^\pm)^{33}_{33}\,.
\eea
Having known these entries, from Eq.\,(\ref{propinverse}), one can derive the full quark propagator matrix
\be\label{prop}
S(K)=\left(
\begin{array}{cc}
G^+(K) & \Xi^-(K) \\
\Xi^+(K) & G^-(K)
\end{array}\right)\;,
\ee
where
\begin{subequations}
\bea\label{fprop}
[G^\pm (K)]^{-1}&=& (G_0^\pm)^{-1}(K)-\phi^\mp G_0^\mp(K)\phi^\pm\;,\\
\Xi^\pm (K)&=&-\,G_0^\mp(K)\phi^\pm G^\pm(K)\;.\label{offprop}
\eea
\end{subequations}
Furthermore, the gap matrix $\Phi^+=-(\Phi^-)^\dagger$ in the gCFL phase has the following form
\bea\label{gapmatrix}
\Phi^\pm=\pm\gamma_5\left(
\begin{array}{ccccccccc}
\sigma_1^\pm & 0 & 0 & 0 & \varphi_3^\pm & 0 & 0 & 0 & \varphi_2^\pm\\
0 & 0 & 0 & \phi_3^\pm & 0 & 0 & 0 & 0 & 0\\
0 & 0 & 0 & 0 & 0 & 0 & \phi_2^\pm & 0 & 0\\
0 & \phi_3^\pm & 0 & 0 & 0 & 0 & 0 & 0 & 0\\
\varphi_3^\pm & 0 & 0 & 0 & \sigma_2^\pm & 0 & 0 & 0 & \varphi_1^\pm\\
0 & 0 & 0 & 0 & 0 & 0 & 0 & \phi_1^\pm & 0\\
0 & 0 & \phi_2^\pm & 0 & 0 & 0 & 0 & 0 & 0\\
0 & 0 & 0 & 0 & 0 & 0 & \phi_1^\pm & 0 & 0\\
\varphi_2^\pm & 0 & 0 & 0 & \varphi_1^\pm & 0 & 0 & 0 & \sigma_3^\pm
\end{array}\right)\;,
\eea
where, $\sigma_i^\pm, \phi_i^\pm$, and $\varphi_i^\pm$ can be written in terms of the singlet and sextet gaps,
\begin{subequations}\label{deltas}
\bea
\left[\Phi^\pm\right]^{11}_{11}\equiv\sigma_1^\pm\equiv 2\Delta_1^{(6)}\,,\nonumber\\
\left[\Phi^\pm\right]^{22}_{22}\equiv\sigma_2^\pm\equiv 2\Delta_2^{(6)}\,,\nonumber\\
\left[\Phi^\pm\right]^{33}_{33}\equiv\sigma_3^\pm\equiv 2\Delta_3^{(6)}\,,
\eea
\bea
\left[\Phi^\pm\right]^{23}_{23}\equiv\left[\Phi^\pm\right]^{32}_{32}\equiv \varphi_1^\pm\equiv \Delta_1^{(6)}+\Delta_1^{({\bar{\bf 3}})}\,,\nonumber\\
\left[\Phi^\pm\right]^{13}_{13}\equiv\left[\Phi^\pm\right]^{31}_{31}\equiv\varphi_2^\pm\equiv \Delta_2^{(6)}+\Delta_2^{({\bar{\bf 3}})}\,,\nonumber\\
\left[\Phi^\pm\right]^{12}_{12}\equiv\left[\Phi^\pm\right]^{21}_{21}\equiv\varphi_3^\pm\equiv \Delta_3^{(6)}+\Delta_3^{({\bar{\bf 3}})}\,,
\eea
\bea
\left[\Phi^\pm\right]^{23}_{32}\equiv\left[\Phi^\pm\right]^{32}_{23}\equiv\phi_1^\pm\equiv \Delta_1^{(6)}-\Delta_1^{({\bar{\bf 3}})}\,,\nonumber\\
\left[\Phi^\pm\right]^{13}_{31}\equiv\left[\Phi^\pm\right]^{31}_{13}\equiv\phi_2^\pm\equiv
\Delta_2^{(6)}-\Delta_2^{({\bar{\bf 3}})}\,,\nonumber\\
\left[\Phi^\pm\right]^{12}_{21}\equiv\left[\Phi^\pm\right]^{21}_{12}\equiv\phi_3^\pm\equiv \Delta_3^{(6)}-\Delta_3^{({\bar{\bf 3}})}\,,
\eea
\end{subequations}
Here, we have assumed that the gaps have real values, $\Delta_i^{({\bar{\bf 3}})}=\Delta_i^{\dagger({\bar{\bf 3}})}$ and $\Delta_i^{(6)}=\Delta_i^{\dagger(6)}$. Employing Eqs.\eeq{comprop} and \eeq{deltas} we can find the explicit form of the propagator Eq.\,(\ref{fprop}) in terms of its color and flavor structure. As already mentioned, the full propagator in the color-flavor space is a $9\times 9$ matrix. There is not a direct way to find all entries of this matrix but to calculate them one by one. Afterwards, having the nonzero entries which are just numbers in the color-flavor space, one can find their inverse $[(G^\pm)]^{fg}_{ij}$ in the Dirac space. First, we have to know the form of the full propagator in Eq.\,(\ref{fprop}) in the color-flavor space
\bea\label{cffullprop}
[(G^\pm)^{-1}]^{fg}_{ij}= \left[(G_0^\pm)^{-1}\right]^{fg}_{ij}-\left[\phi^\mp G_0^\mp(K)\phi^\pm\right]^{fg}_{ij}\;.
\eea
The second term in the right hand side (rhs) has, in principal, the following form
\bea\label{cfrest}
[\,\phi^\mp G_0^\mp\phi^\pm\,]^{fg}_{ij}&=&[\,\phi^\mp\,]^{f\eta}_{ih}\,[\,G_0^\mp\,]
^{\eta\gamma}_{hk}\,[\,\phi^\pm\,]^{\gamma g}_{kj}\;,
\eea
where a sum on the repeated indices $\eta, \gamma, h$, and $k$ is realized. Using Eqs.\eeq{comprop} and \eeq{deltas} one can now find the nonzero components of the propagators. They are
\begin{subequations}\label{ddiag}
\bea
\left[(G^\pm)^{-1}\right]^{11}_{11}&=& \left[(G_0^\pm)^{-1}\right]^{11}_{11} - [\phi^\mp]^{1\eta}_{1h}\,[G_0^\mp]^{\eta\gamma}_{hk}\,[\phi^\pm]^{\gamma 1}_{k 1}\nonumber\\
&=& \left[(G_0^\pm)^{-1}\right]^{11}_{11} - [\phi^\mp]^{11}_{11}\,[G_0^\mp]
^{11}_{11}\,[\phi^\pm]^{11}_{11} - [\phi^\mp]^{12}_{12}\,[G_0^\mp]
^{22}_{22}\,[\phi^\pm]^{21}_{21} - [\phi^\mp]^{13}_{13}\,[G_0^\mp]
^{33}_{33}\,[\phi^\pm]^{31}_{31}\,\,,
\eea
\bea
\left[(G^\pm)^{-1}\right]^{22}_{22}&=& \left[(G_0^\pm)^{-1}\right]^{22}_{22} - [\phi^\mp]^{2\eta}_{2h}\,[G_0^\mp]^{\eta\gamma}_{hk}\,[\phi^\pm]^{\gamma 2}_{k 2}\nonumber\\
&=& \left[(G_0^\pm)^{-1}\right]^{22}_{22} - [\phi^\mp]^{21}_{21}\,[G_0^\mp]
^{11}_{11}\,[\phi^\pm]^{12}_{12} - [\phi^\mp]^{22}_{22}\,[G_0^\mp]
^{22}_{22}\,[\phi^\pm]^{22}_{22} - [\phi^\mp]^{23}_{23}\,[G_0^\mp]
^{33}_{33}\,[\phi^\pm]^{32}_{32}\,\,,
\eea
\bea
\left[(G^\pm)^{-1}\right]^{33}_{33}&=& \left[(G_0^\pm)^{-1}\right]^{33}_{33} - [\phi^\mp]^{3\eta}_{3h}\,[G_0^\mp]^{\eta\gamma}_{hk}\,[\phi^\pm]^{\gamma 3}_{k 3}\nonumber\\
&=& \left[(G_0^\pm)^{-1}\right]^{33}_{33} - [\phi^\mp]^{31}_{31}\,[G_0^\mp]
^{11}_{11}\,[\phi^\pm]^{13}_{13} - [\phi^\mp]^{32}_{32}\,[G_0^\mp]
^{22}_{22}\,[\phi^\pm]^{23}_{23} - [\phi^\mp]^{33}_{33}\,[G_0^\mp]
^{33}_{33}\,[\phi^\pm]^{33}_{33}\,\,,
\eea
\bea
\left[(G^\pm)^{-1}\right]^{12}_{12}&=& - [\phi^\mp]^{1\eta}_{1h}\,[G_0^\mp]^{\eta\gamma}_{hk}\,[\phi^\pm]^{\gamma 2}_{k 2}\nonumber\\
&=& - [\phi^\mp]^{11}_{11}\,[G_0^\mp]
^{11}_{11}\,[\phi^\pm]^{12}_{12} - [\phi^\mp]^{12}_{12}\,[G_0^\mp]
^{22}_{22}\,[\phi^\pm]^{22}_{22} - [\phi^\mp]^{13}_{13}\,[G_0^\mp]
^{33}_{33}\,[\phi^\pm]^{32}_{32}\,\,,
\eea
\bea
\left[(G^\pm)^{-1}\right]^{13}_{13}&=& - [\phi^\mp]^{1\eta}_{1h}\,[G_0^\mp]^{\eta\gamma}_{hk}\,[\phi^\pm]^{\gamma 3}_{k 3}\nonumber\\
&=& - [\phi^\mp]^{11}_{11}\,[G_0^\mp]
^{11}_{11}\,[\phi^\pm]^{13}_{13} - [\phi^\mp]^{12}_{12}\,[G_0^\mp]
^{22}_{22}\,[\phi^\pm]^{23}_{23} - [\phi^\mp]^{13}_{13}\,[G_0^\mp]
^{33}_{33}\,[\phi^\pm]^{33}_{33}\,\,,
\eea
\bea
\left[(G^\pm)^{-1}\right]^{21}_{21}&=& - [\phi^\mp]^{2\eta}_{2h}\,[G_0^\mp]^{\eta\gamma}_{hk}\,[\phi^\pm]^{\gamma 1}_{k 1}\nonumber\\
&=& - [\phi^\mp]^{21}_{21}\,[G_0^\mp]
^{11}_{11}\,[\phi^\pm]^{11}_{11} - [\phi^\mp]^{22}_{22}\,[G_0^\mp]
^{22}_{22}\,[\phi^\pm]^{21}_{21} - [\phi^\mp]^{23}_{23}\,[G_0^\mp]
^{33}_{33}\,[\phi^\pm]^{31}_{31}\,\,,
\eea
\bea
\left[(G^\pm)^{-1}\right]^{23}_{23}&=& - [\phi^\mp]^{2\eta}_{2h}\,[G_0^\mp]^{\eta\gamma}_{hk}\,[\phi^\pm]^{\gamma 3}_{k 3}\nonumber\\
&=& - [\phi^\mp]^{21}_{21}\,[G_0^\mp]
^{11}_{11}\,[\phi^\pm]^{13}_{13} - [\phi^\mp]^{22}_{22}\,[G_0^\mp]
^{22}_{22}\,[\phi^\pm]^{23}_{23} - [\phi^\mp]^{23}_{23}\,[G_0^\mp]
^{33}_{33}\,[\phi^\pm]^{33}_{33}\,\,,
\eea
\bea
\left[(G^\pm)^{-1}\right]^{31}_{31}&=& - [\phi^\mp]^{3\eta}_{3h}\,[G_0^\mp]^{\eta\gamma}_{hk}\,[\phi^\pm]^{\gamma 1}_{k 1}\nonumber\\
&=& - [\phi^\mp]^{31}_{31}\,[G_0^\mp]
^{11}_{11}\,[\phi^\pm]^{11}_{11} - [\phi^\mp]^{32}_{32}\,[G_0^\mp]
^{22}_{22}\,[\phi^\pm]^{21}_{21} - [\phi^\mp]^{33}_{33}\,[G_0^\mp]
^{33}_{33}\,[\phi^\pm]^{31}_{31}\,\,,
\eea
\bea
\left[(G^\pm)^{-1}\right]^{32}_{32}&=& - [\phi^\mp]^{3\eta}_{3h}\,[G_0^\mp]^{\eta\gamma}_{hk}\,[\phi^\pm]^{\gamma 2}_{k 2}\nonumber\\
&=& - [\phi^\mp]^{31}_{31}\,[G_0^\mp]
^{11}_{11}\,[\phi^\pm]^{12}_{12} - [\phi^\mp]^{32}_{32}\,[G_0^\mp]
^{22}_{22}\,[\phi^\pm]^{22}_{22} - [\phi^\mp]^{33}_{33}\,[G_0^\mp]
^{33}_{33}\,[\phi^\pm]^{32}_{32}\,\,,
\eea
\bea
\left[(G^\pm)^{-1}\right]^{22}_{11}&=& \left[(G_0^\pm)^{-1}\right]^{22}_{11} - [\phi^\mp]^{2\eta}_{1h}\,[G_0^\mp]^{\eta\gamma}_{hk}\,[\phi^\pm]^{\gamma 2}_{k 1}\nonumber\\
&=& \left[(G_0^\pm)^{-1}\right]^{22}_{11} - [\phi^\mp]^{21}_{12}\,[G_0^\mp]
^{11}_{22}\,[\phi^\pm]^{12}_{21}\,\,,
\eea
\bea
\left[(G^\pm)^{-1}\right]^{11}_{22}&=& \left[(G_0^\pm)^{-1}\right]^{11}_{22} - [\phi^\mp]^{1\eta}_{2h}\,[G_0^\mp]^{\eta\gamma}_{hk}\,[\phi^\pm]^{\gamma 1}_{k 2}\nonumber\\
&=& \left[(G_0^\pm)^{-1}\right]^{11}_{22} - [\phi^\mp]^{12}_{21}\,[G_0^\mp]
^{22}_{11}\,[\phi^\pm]^{21}_{12}\,\,,
\eea
\bea
\left[(G^\pm)^{-1}\right]^{22}_{33}&=& \left[(G_0^\pm)^{-1}\right]^{22}_{33} - [\phi^\mp]^{2\eta}_{3h}\,[G_0^\mp]^{\eta\gamma}_{hk}\,[\phi^\pm]^{\gamma 2}_{k 3}\nonumber\\
&=& \left[(G_0^\pm)^{-1}\right]^{22}_{33} - [\phi^\mp]^{23}_{32}\,[G_0^\mp]
^{33}_{22}\,[\phi^\pm]^{32}_{23}\,\,,
\eea
\bea
\left[(G^\pm)^{-1}\right]^{33}_{22}&=& \left[(G_0^\pm)^{-1}\right]^{33}_{22} - [\phi^\mp]^{3\eta}_{2h}\,[G_0^\mp]^{\eta\gamma}_{hk}\,[\phi^\pm]^{\gamma 3}_{k 2}\nonumber\\
&=& \left[(G_0^\pm)^{-1}\right]^{33}_{22} - [\phi^\mp]^{32}_{23}\,[G_0^\mp]
^{22}_{33}\,[\phi^\pm]^{23}_{32}\,\,,
\eea
\bea
\left[(G^\pm)^{-1}\right]^{11}_{33}&=& \left[(G_0^\pm)^{-1}\right]^{11}_{33} - [\phi^\mp]^{1\eta}_{3h}\,[G_0^\mp]^{\eta\gamma}_{hk}\,[\phi^\pm]^{\gamma 1}_{k 3}\nonumber\\
&=& \left[(G_0^\pm)^{-1}\right]^{11}_{33} - [\phi^\mp]^{13}_{31}\,[G_0^\mp]
^{33}_{11}\,[\phi^\pm]^{31}_{13}\,\,,
\eea
\bea
\left[(G^\pm)^{-1}\right]^{33}_{11}&=& \left[(G_0^\pm)^{-1}\right]^{33}_{11} - [\phi^\mp]^{3\eta}_{1h}\,[G_0^\mp]^{\eta\gamma}_{hk}\,[\phi^\pm]^{\gamma 3}_{k 1}\nonumber\\
&=& \left[(G_0^\pm)^{-1}\right]^{33}_{11} - [\phi^\mp]^{31}_{13}\,[G_0^\mp]
^{11}_{33}\,[\phi^\pm]^{13}_{31}\,\,.
\eea
\end{subequations}
Inserting $[G_0^{-1}]^{fg}_{ij}$ from \eq{cfprop} and $[\Phi^\pm]^{fg}_{ij}$ from \eq{deltas} into the above equations and inverting the rhs we have
\begin{subequations}\label{diag}
\bea
[G^\pm (K)]^{11}_{11} &=& \sum_{e=\pm}\left[k_0\pm(\mu^{11}_{11}-e k)-\frac{\sigma_1^2}{k_0\mp(\mu^{11}_{11}-e k)}-\frac{\varphi_2^2}{k_0\mp(\mu^{22}_{22}-e k)}-\frac{\varphi_3^2}{k_0\mp(\mu^{33}_{33}-e k)}\right]^{-1}\Lambda^{\pm e}\gamma_0\,\,,
\eea
\bea
[G^\pm (K)]^{22}_{22} &=& \sum_{e=\pm}\left[k_0\pm(\mu^{22}_{22}-e k)-\frac{\varphi_1^2}{k_0\mp(\mu^{11}_{11}-e k)}-\frac{\sigma_2^2}{k_0\mp(\mu^{22}_{22}-e k)}-\frac{\varphi_3^2}{k_0\mp(\mu^{33}_{33}-e k)}\right]^{-1}\Lambda^{\pm e}\gamma_0\,\,,
\eea
\bea
[G^\pm (K)]^{33}_{33} &=& \sum_{e=\pm}\left[k_0\pm(\mu^{33}_{33}-e k)-\frac{\varphi_1^2}{k_0\mp(\mu^{11}_{11}-e k)}-\frac{\varphi_2^2}{k_0\mp(\mu^{22}_{22}-e k)}-\frac{\sigma_3^2}{k_0\mp(\mu^{33}_{33}-e k)}\right]^{-1}\Lambda^{\pm e}\gamma_0\,\,,
\eea
\bea
[G^\pm (K)]^{12}_{12}&=& [G^\pm (K)]^{21}_{21} = -\sum_{e=\pm}\left[\frac{\sigma_1\varphi_3}{k_0\mp(\mu^{11}_{11}-e k)}+\frac{\sigma_2\varphi_3}{k_0\mp(\mu^{22}_{22}-e k)}+\frac{\varphi_1\varphi_2}{k_0\mp(\mu^{33}_{33}-e k)}\right]^{-1}\Lambda^{\pm e}\gamma_0\,\,,
\eea
\bea
[G^\pm (K)]^{13}_{13}&=& [G^\pm (K)]^{31}_{31} = - \sum_{e=\pm}\left[\frac{\sigma_1\varphi_2}{k_0\mp(\mu^{11}_{11}-e k)}+\frac{\varphi_1\varphi_3}{k_0\mp(\mu^{22}_{22}-e k)}+\frac{\sigma_3\varphi_2}{k_0\mp(\mu^{33}_{33}-e k)}\right]^{-1}\Lambda^{\pm e}\gamma_0\,\,,
\eea
\bea
[G^\pm (K)]^{23}_{23}&=& [G^\pm (K)]^{32}_{32} =- \sum_{e=\pm}\left[\frac{\varphi_2\varphi_3}{k_0\mp(\mu^{11}_{11}-e k)}+\frac{\sigma_2\varphi_1}{k_0\mp(\mu^{22}_{22}-e k)}+\frac{\sigma_3\varphi_1}{k_0\mp(\mu^{33}_{33}-e k)}\right]^{-1}\Lambda^{\pm e}\gamma_0\,\,,
\eea
\bea
[G^\pm (K)]^{22}_{11} &=& \sum_{e=\pm}\left[k_0\pm(\mu^{22}_{11}-e k)-\frac{\phi_3^2}{k_0\mp(\mu^{11}_{22}-e k)}\right]^{-1}\Lambda^{\pm e}\gamma_0\,\,,
\eea
\bea
[G^\pm (K)]^{11}_{22} &=& \sum_{e=\pm}\left[k_0\pm(\mu^{11}_{22}-e k)-\frac{\phi_3^2}{k_0\mp(\mu^{22}_{11}-e k)}\right]^{-1}\Lambda^{\pm e}\gamma_0\,\,,
\eea
\bea
[G^\pm (K)]^{22}_{33} &=& \sum_{e=\pm}\left[k_0\pm(\mu^{22}_{33}-e k)-\frac{\phi_1^2}{k_0\mp(\mu^{33}_{22}-e k)}\right]^{-1}\Lambda^{\pm e}\gamma_0\,\,,
\eea
\bea
[G^\pm (K)]^{33}_{22} &=& \sum_{e=\pm}\left[k_0\pm(\mu^{33}_{22}-e k)-\frac{\phi_1^2}{k_0\mp(\mu^{22}_{33}-e k)}\right]^{-1}\Lambda^{\pm e}\gamma_0\,\,,
\eea
\bea
[G^\pm (K)]^{11}_{33} &=& \sum_{e=\pm}\left[k_0\pm(\mu^{11}_{33}-e k)-\frac{\phi_2^2}{k_0\mp(\mu^{33}_{11}-e k)}\right]^{-1}\Lambda^{\pm e}\gamma_0\,\,,
\eea
\bea
[G^\pm (K)]^{33}_{11} &=& \sum_{e=\pm}\left[k_0\pm(\mu^{33}_{11}-e k)-\frac{\phi_2^2}{k_0\mp(\mu^{11}_{33}-e k)}\right]^{-1}\Lambda^{\pm e}\gamma_0\,\,,
\eea
\end{subequations}
where
\bea
\Lambda^\pm=\frac{1}{2}\left(1\pm\gamma_0\,\gamma\cdot\hat{k}\right)\;
\eea
are projectors on the states of positive $\Lambda^+$ or negative $\Lambda^-$ energies. We see that there are new components for the full propagators which do not have any correspondence for the bare propagators, e.g. $[G^\pm (K)]^{12}_{12}$, $[G^\pm (K)]^{13}_{13}$, and etc. Besides, a closer inspection reveals that there are in general three different groups for the propagators. However, as we shall see in the following (Sec.\,\ref{sec2}), in the CFL phase these three groups shrink to two groups as we expect, Ref.\,\cite{3frischke}.

Following the method used to calculate the diagonal components of the quark propagator, \eq{fprop}, we can find the explicit form of the off-diagonal components of the propagator. We write Eq.\,(\ref{offprop}) in terms of its color and flavor structure
\bea
[\,\Xi^\pm\,]^{fg}_{ij}&=&-\,[\,G_0^\mp\,]^{f\eta}_{ih}[\,\phi^\pm\,]^{\eta\gamma}_{hk}[\,G^\pm\,]^{\gamma g}_{kj}\,.
\eea
Having already separated the nonzero components of $[\,G_0^\mp\,]^{fg}_{ij}, [\,G^\pm\,]^{fg}_{ij}$, and $[\,\phi^\pm\,]^{fg}_{ij}$, one can find the nonzero components of the off-diagonal components, one by one checking is needed. They are
\begin{subequations}
\bea
[\,\Xi^\pm\,]^{11}_{11}&=&-\,[\,G_0^\mp\,]^{11}_{11}\bigg([\,\phi^\pm\,]^{11}_{11}[\,G^\pm\,]^{11}_{11}+[\,\phi^\pm\,]^{12}_{12}[\,G^\pm\,]^{21}_{21}+[\,\phi^\pm\,]^{13}_{13}[\,G^\pm\,]^{31}_{31}\bigg)\,,
\eea
\bea
[\,\Xi^\pm\,]^{12}_{12}&=&-\,[\,G_0^\mp\,]^{11}_{11}\bigg([\,\phi^\pm\,]^{11}_{11}[\,G^\pm\,]^{12}_{12}+[\,\phi^\pm\,]^{12}_{12}[\,G^\pm\,]^{22}_{22}+[\,\phi^\pm\,]^{13}_{13}[\,G^\pm\,]^{32}_{32}\bigg)\,,
\eea
\bea
[\,\Xi^\pm\,]^{13}_{13}&=&-\,[\,G_0^\mp\,]^{11}_{11}\bigg([\,\phi^\pm\,]^{11}_{11}[\,G^\pm\,]^{13}_{13}+[\,\phi^\pm\,]^{12}_{12}[\,G^\pm\,]^{23}_{23}+[\,\phi^\pm\,]^{13}_{13}[\,G^\pm\,]^{33}_{33}\bigg)\,,
\eea
\bea
[\,\Xi^\pm\,]^{22}_{22}&=&-\,[\,G_0^\mp\,]^{22}_{22}\bigg([\,\phi^\pm\,]^{21}_{21}[\,G^\pm\,]^{12}_{12}+[\,\phi^\pm\,]^{22}_{22}[\,G^\pm\,]^{22}_{22}+[\,\phi^\pm\,]^{23}_{23}[\,G^\pm\,]^{32}_{32}\bigg)\,,
\eea
\bea
[\,\Xi^\pm\,]^{21}_{21}&=&-\,[\,G_0^\mp\,]^{22}_{22}\bigg([\,\phi^\pm\,]^{21}_{21}[\,G^\pm\,]^{11}_{11}+[\,\phi^\pm\,]^{22}_{22}[\,G^\pm\,]^{21}_{21}+[\,\phi^\pm\,]^{23}_{23}[\,G^\pm\,]^{31}_{31}\bigg)\,,
\eea
\bea
[\,\Xi^\pm\,]^{23}_{23}&=&-\,[\,G_0^\mp\,]^{22}_{22}\bigg([\,\phi^\pm\,]^{21}_{21}[\,G^\pm\,]^{13}_{13}+[\,\phi^\pm\,]^{22}_{22}[\,G^\pm\,]^{23}_{23}+[\,\phi^\pm\,]^{23}_{23}[\,G^\pm\,]^{33}_{33}\bigg)\,,
\eea
\bea
[\,\Xi^\pm\,]^{33}_{33}&=&-\,[\,G_0^\mp\,]^{33}_{33}\bigg([\,\phi^\pm\,]^{31}_{31}[\,G^\pm\,]^{13}_{13}+[\,\phi^\pm\,]^{32}_{32}[\,G^\pm\,]^{23}_{23}+[\,\phi^\pm\,]^{33}_{33}[\,G^\pm\,]^{33}_{33}\bigg)\,,
\eea
\bea
[\,\Xi^\pm\,]^{31}_{31}&=&-\,[\,G_0^\mp\,]^{33}_{33}\bigg([\,\phi^\pm\,]^{31}_{31}[\,G^\pm\,]^{11}_{11}+[\,\phi^\pm\,]^{32}_{32}[\,G^\pm\,]^{21}_{21}+[\,\phi^\pm\,]^{33}_{33}[\,G^\pm\,]^{31}_{31}\bigg)\,,
\eea
\bea
[\,\Xi^\pm\,]^{32}_{32}&=&-\,[\,G_0^\mp\,]^{33}_{33}\bigg([\,\phi^\pm\,]^{31}_{31}[\,G^\pm\,]^{12}_{12}+[\,\phi^\pm\,]^{32}_{32}[\,G^\pm\,]^{22}_{22}+[\,\phi^\pm\,]^{33}_{33}[\,G^\pm\,]^{32}_{32}\bigg)\,,
\eea
\bea
[\,\Xi^\pm\,]^{21}_{12}&=&-\,[\,G_0^\mp\,]^{22}_{11}[\,\phi^\pm\,]^{21}_{12}[\,G^\pm\,]^{11}_{22}\,,
\eea
\bea
[\,\Xi^\pm\,]^{12}_{21}&=&-\,[\,G_0^\mp\,]^{11}_{22}[\,\phi^\pm\,]^{12}_{21}[\,G^\pm\,]^{22}_{11}\,,
\eea
\bea
[\,\Xi^\pm\,]^{31}_{13}&=&-\,[\,G_0^\mp\,]^{33}_{11}[\,\phi^\pm\,]^{31}_{13}[\,G^\pm\,]^{11}_{33}\,,
\eea
\bea
[\,\Xi^\pm\,]^{13}_{31}&=&-\,[\,G_0^\mp\,]^{11}_{33}[\,\phi^\pm\,]^{13}_{31}[\,G^\pm\,]^{33}_{11}\,,
\eea
\bea
[\,\Xi^\pm\,]^{32}_{23}&=&-\,[\,G_0^\mp\,]^{33}_{22}[\,\phi^\pm\,]^{32}_{23}[\,G^\pm\,]^{22}_{33}\,,
\eea
\bea
[\,\Xi^\pm\,]^{23}_{32}&=&-\,[\,G_0^\mp\,]^{22}_{33}[\,\phi^\pm\,]^{23}_{32}[\,G^\pm\,]^{33}_{22}\,,
\eea
\end{subequations}
and inserting the explicit values of the gaps and the propagators we find
\begin{subequations}\label{offdiag}
\bea
[\,\Xi^\pm (K)\,]^{11}_{11} &=& \mp\sum_{e= \pm} \frac{\Lambda^{\mp e}\gamma_5}{k_0\mp(\mu^{22}_{22}-e k)}\nonumber\\
&\times&\Bigg(-\sigma_1\left[k_0\pm(\mu^{11}_{11}-e k)-\frac{\sigma_1^2}{k_0\mp(\mu^{11}_{11}-e k)}-\frac{\varphi_2^2}{k_0\mp(\mu^{22}_{22}-e k)}-\frac{\varphi_3^2}{k_0\mp(\mu^{33}_{33}-e k)}\right]^{-1}\nonumber\\
&+&\varphi_3\left[\frac{\sigma_1\varphi_3}{k_0\mp(\mu^{11}_{11}-e k)}+\frac{\sigma_2\varphi_3}{k_0\mp(\mu^{22}_{22}-e k)}+\frac{\varphi_1\varphi_2}{k_0\mp(\mu^{33}_{33}-e k)}\right]^{-1}\nonumber\\
&+&\varphi_2\left[\frac{\sigma_1\varphi_2}{k_0\mp(\mu^{11}_{11}-e k)}+\frac{\varphi_1\varphi_3}{k_0\mp(\mu^{22}_{22}-e k)}+\frac{\sigma_3\varphi_2}{k_0\mp(\mu^{33}_{33}-e k)}\right]^{-1}\Bigg)\,,
\eea
\bea
[\,\Xi^\pm (K)\,]^{22}_{22} &=& \mp\sum_{e= \pm} \frac{\Lambda^{\mp e}\gamma_5}{k_0\mp(\mu^{22}_{22}-e k)}\nonumber\\
&\times&\Bigg(-\sigma_2\left[k_0\pm(\mu^{22}_{22}-e k)-\frac{\varphi_1^2}{k_0\mp(\mu^{11}_{11}-e k)}-\frac{\sigma_2^2}{k_0\mp(\mu^{22}_{22}-e k)}-\frac{\varphi_3^2}{k_0\mp(\mu^{33}_{33}-e k)}\right]^{-1}\nonumber\\
&+&\varphi_3\left[\frac{\sigma_1\varphi_3}{k_0\mp(\mu^{11}_{11}-e k)}+\frac{\sigma_2\varphi_3}{k_0\mp(\mu^{22}_{22}-e k)}+\frac{\varphi_1\varphi_2}{k_0\mp(\mu^{33}_{33}-e k)}\right]^{-1}\nonumber\\
&+&\varphi_1\left[\frac{\varphi_2\varphi_3}{k_0\mp(\mu^{11}_{11}-e k)}+\frac{\sigma_2\varphi_1}{k_0\mp(\mu^{22}_{22}-e k)}+\frac{\sigma_3\varphi_1}{k_0\mp(\mu^{33}_{33}-e k)}\right]^{-1}\Bigg)\,,
\eea
\bea
[\,\Xi^\pm (K)\,]^{33}_{33} &=& \mp\sum_{e= \pm} \frac{\Lambda^{\mp e}\gamma_5}{k_0\mp(\mu^{33}_{33}-e k)}\nonumber\\
&\times&\Bigg(-\sigma_3\left[k_0\pm(\mu^{33}_{33}-e k)-\frac{\varphi_1^2}{k_0\mp(\mu^{11}_{11}-e k)}-\frac{\varphi_2^2}{k_0\mp(\mu^{22}_{22}-e k)}-\frac{\sigma_3^2}{k_0\mp(\mu^{33}_{33}-e k)}\right]^{-1}\nonumber\\
&+&\varphi_2\left[\frac{\sigma_1\varphi_2}{k_0\mp(\mu^{11}_{11}-e k)}+\frac{\varphi_1\varphi_3}{k_0\mp(\mu^{22}_{22}-e k)}+\frac{\sigma_3\varphi_2}{k_0\mp(\mu^{33}_{33}-e k)}\right]^{-1}\nonumber\\
&+&\varphi_1\left[\frac{\varphi_2\varphi_3}{k_0\mp(\mu^{11}_{11}-e k)}+\frac{\sigma_2\varphi_1}{k_0\mp(\mu^{22}_{22}-e k)}+\frac{\sigma_3\varphi_1}{k_0\mp(\mu^{33}_{33}-e k)}\right]^{-1}\Bigg)\,,
\eea
\bea
[\,\Xi^\pm (K)\,]^{12}_{12} &=& \mp\sum_{e= \pm} \frac{\Lambda^{\mp e}\gamma_5}{k_0\mp(\mu^{11}_{11}-e k)}\nonumber\\
&\times&\Bigg(-\varphi_3\left[k_0\pm(\mu^{22}_{22}-e k)-\frac{\varphi_1^2}{k_0\mp(\mu^{11}_{11}-e k)}-\frac{\sigma_2^2}{k_0\mp(\mu^{22}_{22}-e k)}-\frac{\varphi_3^2}{k_0\mp(\mu^{33}_{33}-e k)}\right]^{-1}\nonumber\\
&+&\sigma_1\left[\frac{\sigma_1\varphi_3}{k_0\mp(\mu^{11}_{11}-e k)}+\frac{\sigma_2\varphi_3}{k_0\mp(\mu^{22}_{22}-e k)}+\frac{\varphi_1\varphi_2}{k_0\mp(\mu^{33}_{33}-e k)}\right]^{-1}\nonumber\\
&+&\varphi_2\left[\frac{\varphi_2\varphi_3}{k_0\mp(\mu^{11}_{11}-e k)}+\frac{\sigma_2\varphi_1}{k_0\mp(\mu^{22}_{22}-e k)}+\frac{\sigma_3\varphi_1}{k_0\mp(\mu^{33}_{33}-e k)}\right]^{-1}\Bigg)\,,
\eea
\bea
[\,\Xi^\pm (K)\,]^{21}_{21} &=& \mp\sum_{e= \pm} \frac{\Lambda^{\mp e}\gamma_5}{k_0\mp(\mu^{22}_{22}-e k)}\nonumber\\
&\times&\Bigg(-\varphi_3\left[k_0\pm(\mu^{11}_{11}-e k)-\frac{\sigma_1^2}{k_0\mp(\mu^{11}_{11}-e k)}-\frac{\varphi_2^2}{k_0\mp(\mu^{22}_{22}-e k)}-\frac{\varphi_3^2}{k_0\mp(\mu^{33}_{33}-e k)}\right]^{-1}\nonumber\\
&+&\sigma_2\left[\frac{\sigma_1\varphi_3}{k_0\mp(\mu^{11}_{11}-e k)}+\frac{\sigma_2\varphi_3}{k_0\mp(\mu^{22}_{22}-e k)}+\frac{\varphi_1\varphi_2}{k_0\mp(\mu^{33}_{33}-e k)}\right]^{-1}\nonumber\\
&+&\varphi_1\left[\frac{\sigma_1\varphi_2}{k_0\mp(\mu^{11}_{11}-e k)}+\frac{\varphi_1\varphi_3}{k_0\mp(\mu^{22}_{22}-e k)}+\frac{\sigma_3\varphi_2}{k_0\mp(\mu^{33}_{33}-e k)}\right]^{-1}\Bigg)\,,
\eea
\bea
[\,\Xi^\pm (K)\,]^{13}_{13} &=& \mp\sum_{e= \pm} \frac{\Lambda^{\mp e}\gamma_5}{k_0\mp(\mu^{11}_{11}-e k)}\nonumber\\
&\times&\Bigg(-\varphi_2\left[k_0\pm(\mu^{33}_{33}-e k)-\frac{\varphi_1^2}{k_0\mp(\mu^{11}_{11}-e k)}-\frac{\varphi_2^2}{k_0\mp(\mu^{22}_{22}-e k)}-\frac{\sigma_3^2}{k_0\mp(\mu^{33}_{33}-e k)}\right]^{-1}\nonumber\\
&+&\sigma_1\left[\frac{\sigma_1\varphi_2}{k_0\mp(\mu^{11}_{11}-e k)}+\frac{\varphi_1\varphi_3}{k_0\mp(\mu^{22}_{22}-e k)}+\frac{\sigma_3\varphi_2}{k_0\mp(\mu^{33}_{33}-e k)}\right]^{-1}\nonumber\\
&+&\varphi_3\left[\frac{\varphi_2\varphi_3}{k_0\mp(\mu^{11}_{11}-e k)}+\frac{\sigma_2\varphi_1}{k_0\mp(\mu^{22}_{22}-e k)}+\frac{\sigma_3\varphi_1}{k_0\mp(\mu^{33}_{33}-e k)}\right]^{-1}\Bigg)\,,
\eea
\bea
[\,\Xi^\pm (K)\,]^{31}_{31} &=& \mp\sum_{e= \pm}\frac{\Lambda^{\mp e} \gamma_5}{k_0\mp(\mu^{22}_{22}-e k)}\nonumber\\
&\times&\Bigg(-\varphi_2\left[k_0\pm(\mu^{11}_{11}-e k)-\frac{\sigma_1^2}{k_0\mp(\mu^{11}_{11}-e k)}-\frac{\varphi_2^2}{k_0\mp(\mu^{22}_{22}-e k)}-\frac{\varphi_3^2}{k_0\mp(\mu^{33}_{33}-e k)}\right]^{-1}\nonumber\\
&+&\sigma_3\left[\frac{\sigma_1\varphi_2}{k_0\mp(\mu^{11}_{11}-e k)}+\frac{\varphi_1\varphi_3}{k_0\mp(\mu^{22}_{22}-e k)}+\frac{\sigma_3\varphi_2}{k_0\mp(\mu^{33}_{33}-e k)}\right]^{-1}\nonumber\\
&+&\varphi_1\left[\frac{\sigma_1\varphi_3}{k_0\mp(\mu^{11}_{11}-e k)}+\frac{\sigma_2\varphi_3}{k_0\mp(\mu^{22}_{22}-e k)}+\frac{\varphi_1\varphi_2}{k_0\mp(\mu^{33}_{33}-e k)}\right]^{-1}\Bigg)\,,
\eea
\bea
[\,\Xi^\pm (K)\,]^{23}_{23} &=&\mp\sum_{e= \pm}\frac{\Lambda^{\mp e}\gamma_5}{k_0\mp(\mu^{22}_{22}-e k)}\nonumber\\
&\times&\Bigg(-\varphi_1\left[k_0\pm(\mu^{33}_{33}-e k)-\frac{\varphi_1^2}{k_0\mp(\mu^{11}_{11}-e k)}-\frac{\varphi_2^2}{k_0\mp(\mu^{22}_{22}-e k)}-\frac{\sigma_3^2}{k_0\mp(\mu^{33}_{33}-e k)}\right]^{-1}\nonumber\\
&+&\sigma_2\left[\frac{\varphi_2\varphi_3}{k_0\mp(\mu^{11}_{11}-e k)}+\frac{\sigma_2\varphi_1}{k_0\mp(\mu^{22}_{22}-e k)}+\frac{\sigma_3\varphi_1}{k_0\mp(\mu^{33}_{33}-e k)}\right]^{-1}\nonumber\\
&+&\varphi_3\left[\frac{\sigma_1\varphi_2}{k_0\mp(\mu^{11}_{11}-e k)}+\frac{\varphi_1\varphi_3}{k_0\mp(\mu^{22}_{22}-e k)}+\frac{\sigma_3\varphi_2}{k_0\mp(\mu^{33}_{33}-e k)}\right]^{-1}\Bigg)\,,
\eea
\bea
[\,\Xi^\pm (K)\,]^{32}_{32} &=& \mp\sum_{e= \pm} \frac{\Lambda^{\mp e}\gamma_5}{k_0\mp(\mu^{33}_{33}-e k)}\nonumber\\
&\times&\Bigg(-\varphi_1\left[k_0\pm(\mu^{22}_{22}-e k)-\frac{\varphi_1^2}{k_0\mp(\mu^{11}_{11}-e k)}-\frac{\sigma_2^2}{k_0\mp(\mu^{22}_{22}-e k)}-\frac{\varphi_3^2}{k_0\mp(\mu^{33}_{33}-e k)}\right]^{-1}\nonumber\\
&+&\sigma_3\left[\frac{\varphi_2\varphi_3}{k_0\mp(\mu^{11}_{11}-e k)}+\frac{\sigma_2\varphi_1}{k_0\mp(\mu^{22}_{22}-e k)}+\frac{\sigma_3\varphi_1}{k_0\mp(\mu^{33}_{33}-e k)}\right]^{-1}\nonumber\\
&+&\varphi_2\left[\frac{\sigma_1\varphi_3}{k_0\mp(\mu^{11}_{11}-e k)}+\frac{\sigma_2\varphi_3}{k_0\mp(\mu^{22}_{22}-e k)}+\frac{\varphi_1\varphi_2}{k_0\mp(\mu^{33}_{33}-e k)}\right]^{-1}\Bigg)\,,
\eea
\bea
[\,\Xi^\pm (K)\,]^{12}_{21} &=& \pm\sum_{e= \pm}\frac{\Lambda^{\mp e}\gamma_5\,\phi_3}{k_0\mp(\mu^{11}_{22}-e k)}\left[k_0\pm(\mu^{22}_{11}-e k)-\frac{\phi_3^2}{k_0\mp(\mu^{11}_{22}-e k)}\right]^{-1}\,,
\eea
\bea
[\,\Xi^\pm (K)\,]^{21}_{12} &=& \pm\sum_{e= \pm}\frac{\Lambda^{\mp e} \gamma_5 \,\phi_3}{k_0\mp(\mu^{22}_{11}-e k)}\left[k_0\pm(\mu^{11}_{22}-e k)-\frac{\phi_3^2}{k_0\mp(\mu^{22}_{11}-e k)}\right]^{-1}\,,
\eea
\bea
[\,\Xi^\pm (K)\,]^{13}_{31} &=& \pm\sum_{e= \pm}\frac{\Lambda^{\mp e} \gamma_5\,\phi_2}{k_0\mp(\mu^{11}_{33}-e k)}\left[k_0\pm(\mu^{33}_{11}-e k)-\frac{\phi_2^2}{k_0\mp(\mu^{11}_{33}-e k)}\right]^{-1}\,,
\eea
\bea
[\,\Xi^\pm (K)\,]^{31}_{13} &=& \pm\sum_{e= \pm}\frac{\Lambda^{\mp e} \gamma_5 \,\phi_2}{k_0\mp(\mu^{33}_{11}-e k)}\left[k_0\pm(\mu^{11}_{33}-e k)-\frac{\phi_2^2}{k_0\mp(\mu^{33}_{11}-e k)}\right]^{-1}\,,
\eea
\bea
[\,\Xi^\pm (K)\,]^{23}_{32} &=& \pm\sum_{e= \pm}\frac{\Lambda^{\mp e} \gamma_5 \,\phi_1}{k_0\mp(\mu^{22}_{33}-e k)}\left[k_0\pm(\mu^{33}_{22}-e k)-\frac{\phi_1^2}{k_0\mp(\mu^{22}_{33}-e k)}\right]^{-1}\,,
\eea
\bea
[\,\Xi^\pm (K)\,]^{32}_{23} &=& \pm\sum_{e= \pm}\frac{\Lambda^{\mp e} \gamma_5\phi_1}{k_0\mp(\mu^{33}_{22}-e k)}\left[k_0\pm(\mu^{22}_{33}-e k)-\frac{\phi_1^2}{k_0\mp(\mu^{33}_{22}-e k)}\right]^{-1}\,.
\eea
\end{subequations}
These results can be generalized to the case we have an external magnetic field in the system. Then, one can use them to calculate the properties of the magnetic-CFL phase. In the next section we present the diagonal and the off-diagonal propagators in the CFL phase. We use them in this paper when we calculate the explicit form of the gluon self-energies in the CFL phase. Further details are explained in the following.

\section{Quark propagators in CFL phase}\label{sec2}

We now know that the CFL phase is the dominant phase of CSC at very large densities. Since quark matter in the color superconducting phase constitutes the core of neutron stars, the properties of the matter in the CFL phase should be studied in great details. The result of this section can be used to yield further information about the properties of the matter in the CFL phase.

The CFL phase is characterized by equal values for the gaps
\bea\label{allgaps}
\sigma_1=\sigma_2\hspace{-.1cm}&=&\hspace{-.1cm}\sigma_3=\sigma\,,\nonumber\\
\varphi_1=\varphi_2\hspace{-.1cm}&=&\hspace{-.1cm}\varphi_3=\varphi\,,\nonumber\\
\phi_1=\phi_2\hspace{-.1cm}&=&\hspace{-.1cm}\phi_3=\phi\,.
\eea
In addition, when the strange quark mass $m_s$ is not so large as to disrupt any Cooper pairs, $\mu_e$, $\mu_3$ and $\mu_8$ satisfying electric and color charge neutrality conditions are connected via
\bea\label{mu3mu8}
\mu_3&=&\mu_e\;,\nonumber\\
\mu_8&=&-\frac{m_s^2}{2\mu}+\frac{\mu_e}{2}\;.
\eea
On the other hand, we know that the CFL phase is electrically neutral $N_e=0$, even in the presence of an electron chemical potential, cf. Refs.\,\cite{alfraj, rajwilc}. Since
\bea
N_e=\frac{\mu_e^3}{3\pi^2}\;,
\eea
from \eq{mu3mu8} we have
\bea\label{allmu}
\mu_3&=&0\;,\nonumber\\
\mu_8&=&-\frac{m_s^2}{2\mu}\;.
\eea
These relations simplify the chemical potentials we have in \eq{chemical} to
\begin{subequations}\label{mus}
\bea
\mu_{11}^{11}= \mu_{22}^{22}= \mu_{33}^{33}&=& \mu_{11}^{22}= \mu_{22}^{11}= \mu - \frac{m_s^2}{2\mu}\;,\\
\mu_{11}^{33}= \mu_{22}^{33}&=& \mu - \frac{2m_s^2}{3\mu}\;,\\
\mu_{33}^{11}= \mu_{33}^{11}&=& \mu + \frac{m_s^2}{3\mu}\;.
\eea
\end{subequations}
Now we can replace them in the diagonal and the off-diagonal component of the full propagators, \eqq{diag}{offdiag}, to find the respective values for the CFL phase
\begin{subequations}\label{cflprop}
\bea
[G^\pm(K)]^{11}_{11}=[G^\pm(K)]^{22}_{22}=[G^\pm(K)]^{33}_{33}=\sum_{e=\pm}\left\{\frac{k_0\mp(\mu^{11}_{11}-e k)}{\left[k_0^2-(\mu^{11}_{11}-e k)^2\right] - 2\varphi^2 - \sigma^2}\right\}\Lambda^{\pm e}\gamma_0\;,
\eea
\bea
[G^\pm(K)]^{12}_{12}=\hspace{-.1cm}[G^\pm(K)]^{21}_{21}=[G^\pm(K)]^{13}_{13}=[G^\pm(K)]^{31}_{31}=[G^\pm(K)]^{32}_{32}=-\sum_{e=\pm}\left\{\frac{k_0\mp(\mu^{11}_{11}-e k)}{\varphi^2+2\,\sigma\,\varphi}\right\}\Lambda^{\pm e}\gamma_0\;,
\eea
\bea
[G^\pm(K)]^{11}_{22}=[G^\pm(K)]^{22}_{11}=\sum_{e=\pm}\left\{\frac{k_0\mp(\mu^{11}_{22}-e k)}{\left[k_0^2-(\mu^{11}_{22}-e k)^2\right]-\phi^2}\right\}\Lambda^{\pm e}\gamma_0\;,
\eea
\bea
[G^\pm(K)]^{11}_{33}=[G^\pm(K)]^{22}_{33}=\sum_{e=\pm}\left\{\frac{k_0\mp(\mu^{33}_{11}-e k)}{\left[k_0\mp(\mu^{33}_{11}-e k)\right]\left[k_0\pm(\mu^{11}_{33}-e k)\right]-\phi^2}\right\}\Lambda^{\pm e}\gamma_0\;,
\eea
\bea
[G^\pm(K)]^{33}_{11}=[G^\pm(K)]^{33}_{22}=\sum_{e=\pm}\left\{\frac{k_0\mp(\mu^{11}_{33}-e k)}{\left[k_0\pm(\mu^{33}_{11}-e k)\right]\left[k_0\mp(\mu^{11}_{33}-e k)\right]-\phi^2}\right\}\Lambda^{\pm e}\gamma_0\;.
\eea
\end{subequations}
and from Eqs.\eeq{offdiag} the off-diagonal components read
\begin{subequations}\label{cfloffprop}
\bea
[\Xi^\pm(K)]^{11}_{11}=[\Xi^\pm(K)]^{22}_{22}=[\Xi^\pm(K)]^{33}_{33}= \pm\sum_{e=\pm}\Lambda^{\mp e}\gamma_5\,\Bigg(\frac{\sigma}{k_0^2-(\mu^{11}_{11}+e k)^2-2\,\varphi^2-\sigma^2}-\frac{2}{2\,\sigma+\varphi}\Bigg)
\eea
\bea
[\Xi^\pm(K)]^{12}_{12}&=&[\Xi^\pm(K)]^{21}_{21}=[\Xi^\pm(K)]^{13}_{13}=[\Xi^\pm(K)]^{31}_{31}\nonumber\\
&=&[\Xi^\pm(K)]^{23}_{23}=[\Xi^\pm(K)]^{32}_{32}= \pm\sum_{e=\pm}\Lambda^{\mp e}\gamma_5\,\Bigg(\frac{\varphi}{k_0^2-(\mu^{11}_{11}+e k)^2-2\,\varphi^2-\sigma^2}-\frac{\sigma+\varphi}{2\,\sigma\varphi+\varphi^2}\Bigg)
\eea
\bea
[\Xi^\pm(K)]^{12}_{21}&=&[\Xi^\pm(K)]^{21}_{12}=\pm\sum_{e=\pm}\frac{\Lambda^{\mp e}\gamma_5\,\phi}{k_0^2-(\mu^{11}_{22}-e k)^2-\phi^2}
\eea
\bea
[\Xi^\pm(K)]^{13}_{31}&=&[\Xi^\pm(K)]^{23}_{32}=\pm\sum_{e=\pm}\frac{\Lambda^{\mp e}\gamma_5\,\phi}{\left[k_0\pm(\mu^{33}_{11}-e k)\right]\left[k_0\mp(\mu^{11}_{33}-e k)\right]-\phi^2}
\eea
\bea
[\Xi^\pm(K)]^{31}_{13}&=&[\Xi^\pm(K)]^{32}_{23}=\pm\sum_{e=\pm}\frac{\Lambda^{\mp e}\gamma_5\,\phi}{\left[k_0\mp(\mu^{33}_{11}-e k)\right]\left[k_0\pm(\mu^{11}_{33}-e k)\right]-\phi^2}
\eea
\end{subequations}

In the following, we take one step further and calculate the gluon self-energies in the gapless as well as the ordinary CFL phase. The analytical results, which are given in the following, are very complicated, therefore to find the physical quantities like the spectral densities, the screening mass, and etc, one should employ them in the numerical calculations.

\section{Gluon self-energies in gapless CFL phase}\label{neutralcfl}

At one-loop approximation and at zero temperature, the gluon-self energy is dominated by the quark loop \cite{2frischke,2sc} and has the following form
\bea\label{self-energy}
\Pi^{\mu\nu}_{ab}(P)=\frac{1}{2}g^2\frac{T}{V}\sum_K {\rm Tr}_{s, c, f, NG}\left[\,\hat{\Gamma}^\mu_a\,S(K)\,\hat{\Gamma}^\nu_b\,S(K-P)\right]\,,
\eea
where $P$ is the quark four-momentum, $K$ is the gluon four-momentum, and the trace is over spinor, color, flavor and Nambu-Gorkov spaces. The vertices are defined through
\bea\label{gammas}
\hat{\Gamma}^\mu_a\equiv\left(
\begin{array}{cc}
\Gamma^\mu_a & 0 \\
0 & \bar{\Gamma}^\mu_a
\end{array}\right)
\equiv\left(
\begin{array}{cc}
\gamma^\mu T_a & 0 \\
0 & -\gamma^\mu T^T_a
\end{array}\right)\;.
\eea
In order to evaluate the self-energy, first of all, we have to perform the traces. The trace over Nambu-Gorkov space is trivial and can be done easily. The trace over the color and the flavor spaces in the (g)CFL phase, however, is very complicated because these two spaces are locked to each other. Nevertheless, we will make use of an elementary argument which, to a large extent, helps us to simplify the calculations. Afterwards, we have to evaluate the trace over the spinor space. We leave this part to Ref.\,\cite{future}, where we use the self-energies to find the spectral densities. For the latter, we have to introduce the mixed representations. This requires a sum over Matsubara frequencies; the details are given in the last section of this article.

\subsection{Trace over Nambu-Gorkov space}

Starting from \eq{self-energy} and employing \eq{gammas}, the trace over the Nambu-Gorkov space can be evaluated, The result is
\bea
\Pi^{\mu\nu}_{ab}(P)&=&\frac{g^2 T}{2V}\sum_K {\rm Tr}_{s, c, f}\left[\,\Gamma^\mu_a \,G^+(K)\,\Gamma^\nu_b \,G^+(K-P)+ \bar{\Gamma}^\mu_a\, G^-(K)\,\bar{\Gamma}^\nu_b \,G^-(K-P) \right. \nonumber \\
&+&\left. \Gamma^\mu_a\,\Xi^-(K)\,\bar{\Gamma}^\nu_b\,\Xi^+(K-P)+\bar{\Gamma}^\mu_a\,\Xi^+(K)\,\Gamma^\nu_b\,\Xi^-(K-P)\,\right]\;.
\eea
where $G^\pm$ and $\Xi^\pm$ are calculated in \eqq{diag}{offdiag}. Now, we have to find the trace over the color and flavor spaces.

\subsection{Trace over color and flavor spaces}

Since in the (g)CFL phase the color and the flavor spaces are locked, it is not possible to find the trace over these spaces separately. We know that when the neutrality condition is not enforced, because all chemical potentials are equal, $\mu_{ij}^{fg}=\mu$, the quark propagators all have the same form. Therefore, one can use a set of projectors introduced in Ref.\,\cite{shovwij} to do the trace. In contrast, including the neutrality condition, as we presented, the propagators find different forms for different colors and flavors. Therefore it is a very tedious work to follow the method used in Ref.\,\cite{3frischke}. Instead, in the following, we present an old fashion method.

We write the self-energies in terms of their color and flavor indices
\bea
[\Pi^{\mu\nu}_{ab}(P)]^{fg}_{ij}&=&\frac{g^2T}{2V}\sum_K {\rm Tr}_{s, c, f}\left\{\,\left[\,\Gamma^\mu_a \,\right]_{ik} \,\left[\,G^+(K)\,\right]^{fh}_{kl}\,\left[\,\Gamma^\nu_b\,\right]_{lm} \,\left[\,G^+(K-P)\,\right]^{hg}_{mj}\right.\nonumber\\
&+&\left.\left[\,\bar{\Gamma}^\mu_a\,\right]_{ik}\, \left[\,G^-(K)\,\right]^{fh}_{kl}\,\left[\,\bar{\Gamma}^\nu_b\,\right]_{lm} \,\left[\,G^-(K-P)\,\right]^{hg}_{mj} \right. \nonumber \\
&+&\left.\left[\,\Gamma^\mu_a\,\right]_{ik}\,\left[\,\Xi^-(K)\,\right]^{fh}_{kl}\,\left[\,\bar{\Gamma}^\nu_b\,\right]_{lm}\,\left[\,\Xi^+(K-P)\,\right]^{hg}_{mj}\right.\nonumber\\
&+&\left.\left[\,\bar{\Gamma}^\mu_a\,\right]_{ik}\,\left[\,\Xi^+(K)\,\right]^{fh}_{kl}\,\left[\,\Gamma^\nu_b\,\right]_{lm}\,\left[\,\Xi^-(K-P)\,\right]^{hg}_{mj}\,\right\}\;.
\eea
Now we can find the trace over the color, $i, j$ and the flavor $f, g$ spaces for each $a, b$, so that, $a,b=1,\dots,8$. The color and the flavor indices run from 1 to 3. We do one by one checking. We admit that this is a very boring method, but we do not have another choice. The results are
\begin{subequations}
\bea
\Pi^{\mu\nu}_{11}(P)&=& \Pi^{\mu\nu}_{22}(P) = \frac{g^2T}{8V}\sum_K {\rm Tr}_{s}\left[\right.\nonumber\\
&+&\left.\gamma^\mu\,G^{+11}_{11}(K)\,\gamma^\nu\,G^{+11}_{22}(K-P)+\gamma^\mu\,G^{+22}_{11}(K)\,\gamma^\nu\,G^{+22}_{22}(K-P)+\gamma^\mu\,G^{+33}_{11}(K)\,\gamma^\nu\,G^{+33}_{22}(K-P)\right.\nonumber\\
&+&\left.\gamma^\mu\,G^{+11}_{22}(K)\,\gamma^\nu\,G^{+11}_{11}(K-P)+\gamma^\mu\,G^{+22}_{22}(K)\,\gamma^\nu\,G^{+22}_{11}(K-P)+\gamma^\mu\,G^{+33}_{22}(K)\,\gamma^\nu\,G^{+33}_{11}(K-P)\right.\nonumber\\
&+&\left.\gamma^\mu\,G^{-11}_{11}(K)\,\gamma^\nu\,G^{-11}_{22}(K-P)+\gamma^\mu\,G^{-22}_{11}(K)\,\gamma^\nu\,G^{-22}_{22}(K-P)+\gamma^\mu\,G^{-33}_{11}(K)\,\gamma^\nu\,G^{-33}_{22}(K-P)\right.\nonumber\\
&+&\left.\gamma^\mu\,G^{-11}_{22}(K)\,\gamma^\nu\,G^{-11}_{11}(K-P)+\gamma^\mu\,G^{-22}_{22}(K)\,\gamma^\nu\,G^{-22}_{11}(K-P)+\gamma^\mu\,G^{-33}_{22}(K)\,\gamma^\nu\,G^{-33}_{11}(K-P)\right.\nonumber\\
&+&\left. \gamma^\mu\,\Xi^{-12}_{21}(K)\,\gamma^\nu\,\Xi^{+21}_{21}(K-P)+\gamma^\mu\,\Xi^{-21}_{21}(K)\,\gamma^\nu\,\Xi^{+12}_{21}(K-P) \right.\nonumber\\
&+& \left. \gamma^\mu\,\Xi^{-12}_{12}(K)\,\gamma^\nu\,\Xi^{+21}_{12}(K-P)+\gamma^\mu\,\Xi^{-21}_{12}(K)\,\gamma^\nu\,\Xi^{+12}_{12}(K-P)\right.\nonumber\\
&+&\left. \gamma^\mu\,\Xi^{+12}_{21}(K)\,\gamma^\nu\,\Xi^{-21}_{21}(K-P)+\gamma^\mu\,\Xi^{+21}_{21}(K)\,\gamma^\nu\,\Xi^{-12}_{21}(K-P)\right.\nonumber\\
&+&\left. \gamma^\mu\,\Xi^{+12}_{12}(K)\,\gamma^\nu\,\Xi^{-21}_{12}(K-P)+\gamma^\mu\,\Xi^{+21}_{12}(K)\,\gamma^\nu\,\Xi^{-12}_{12}(K-P)\,
\right]\;,
\eea
\bea
\Pi^{\mu\nu}_{33}(P)&=&\frac{g^2T}{8V}\sum_K {\rm Tr}_{s}\left[\right.\nonumber\\
&+&\left.\gamma^\mu\,G^{+11}_{11}(K)\,\gamma^\nu\,G^{+11}_{11}(K-P)+\gamma^\mu\,G^{+22}_{11}(K)\,\gamma^\nu\,G^{+22}_{11}(K-P)+\gamma^\mu\,G^{+33}_{11}(K)\,\gamma^\nu\,G^{+33}_{11}(K-P)\right.\nonumber\\
&+&\left.\gamma^\mu\,G^{+11}_{22}(K)\,\gamma^\nu\,G^{+11}_{22}(K-P)+\gamma^\mu\,G^{+22}_{22}(K)\,\gamma^\nu\,G^{+22}_{22}(K-P)+\gamma^\mu\,G^{+33}_{22}(K)\,\gamma^\nu\,G^{+33}_{22}(K-P)\right.\nonumber\\
&-&\left.\gamma^\mu\,G^{+12}_{12}(K)\,\gamma^\nu\,G^{+21}_{21}(K-P)-\gamma^\mu\,G^{+21}_{21}(K)\,\gamma^\nu\,G^{+12}_{12}(K-P)\right.\nonumber\\
&+&\left.\gamma^\mu\,G^{-11}_{11}(K)\,\gamma^\nu\,G^{-11}_{11}(K-P)+\gamma^\mu\,G^{-22}_{11}(K)\,\gamma^\nu\,G^{-22}_{11}(K-P)+\gamma^\mu\,G^{-33}_{11}(K)\,\gamma^\nu\,G^{-33}_{11}(K-P)\right.\nonumber\\
&+&\left.\gamma^\mu\,G^{-11}_{22}(K)\,\gamma^\nu\,G^{-11}_{22}(K-P)+\gamma^\mu\,G^{-22}_{22}(K)\,\gamma^\nu\,G^{-22}_{22}(K-P)+\gamma^\mu\,G^{-33}_{22}(K)\,\gamma^\nu\,G^{-33}_{22}(K-P)\right.\nonumber\\
&-&\left.\gamma^\mu\,G^{-12}_{12}(K)\,\gamma^\nu\,G^{-21}_{21}(K-P)-\gamma^\mu\,G^{-21}_{21}(K)\,\gamma^\nu\,G^{-12}_{12}(K-P)\right.\nonumber\\
&+&\left. \gamma^\mu\,\Xi^{-11}_{11}(K)\,\gamma^\nu\,\Xi^{+11}_{11}(K-P)+\gamma^\mu\,\Xi^{-22}_{22}(K)\,\gamma^\nu\,\Xi^{+22}_{22}(K-P)\right.\nonumber\\
&-&\left. \gamma^\mu\,\Xi^{-12}_{12}(K)\,\gamma^\nu\,\Xi^{+21}_{21}(K-P)-\gamma^\mu\,\Xi^{-21}_{12}(K)\,\gamma^\nu\,\Xi^{+12}_{21}(K-P) \right.\nonumber\\
&-& \left. \gamma^\mu\,\Xi^{-12}_{21}(K)\,\gamma^\nu\,\Xi^{+21}_{12}(K-P)-\gamma^\mu\,\Xi^{-21}_{21}(K)\,\gamma^\nu\,\Xi^{+12}_{12}(K-P)\right.\nonumber\\
&+&\left. \gamma^\mu\,\Xi^{+11}_{11}(K)\,\gamma^\nu\,\Xi^{-11}_{11}(K-P)+\gamma^\mu\,\Xi^{+22}_{22}(K)\,\gamma^\nu\,\Xi^{-22}_{22}(K-P)\right.\nonumber\\
&-&\left. \gamma^\mu\,\Xi^{+12}_{12}(K)\,\gamma^\nu\,\Xi^{-21}_{21}(K-P)-\gamma^\mu\,\Xi^{+21}_{12}(K)\,\gamma^\nu\,\Xi^{-12}_{21}(K-P)\right.\nonumber\\
&-&\left. \gamma^\mu\,\Xi^{+12}_{21}(K)\,\gamma^\nu\,\Xi^{-21}_{12}(K-P)-\gamma^\mu\,\Xi^{+21}_{21}(K)\,\gamma^\nu\,\Xi^{-12}_{12}(K-P)\,
\right]\;,
\eea
\bea
\Pi^{\mu\nu}_{44}(P)&=& \Pi^{\mu\nu}_{55}(P) = \frac{g^2T}{8V}\sum_K {\rm Tr}_{s}\left[\right.\nonumber\\
&+&\left.\gamma^\mu\,G^{+11}_{11}(K)\,\gamma^\nu\,G^{+11}_{33}(K-P)+\gamma^\mu\,G^{+22}_{11}(K)\,\gamma^\nu\,G^{+22}_{33}(K-P)+\gamma^\mu\,G^{+33}_{11}(K)\,\gamma^\nu\,G^{+33}_{33}(K-P)\right.\nonumber\\
&+&\left.\gamma^\mu\,G^{+11}_{33}(K)\,\gamma^\nu\,G^{+11}_{11}(K-P)+\gamma^\mu\,G^{+22}_{33}(K)\,\gamma^\nu\,G^{+22}_{11}(K-P)+\gamma^\mu\,G^{+33}_{33}(K)\,\gamma^\nu\,G^{+33}_{11}(K-P)\right.\nonumber\\
&+&\left.\gamma^\mu\,G^{-11}_{11}(K)\,\gamma^\nu\,G^{-11}_{33}(K-P)+\gamma^\mu\,G^{-22}_{11}(K)\,\gamma^\nu\,G^{-22}_{33}(K-P)+\gamma^\mu\,G^{-33}_{11}(K)\,\gamma^\nu\,G^{-33}_{33}(K-P)\right.\nonumber\\
&+&\left.\gamma^\mu\,G^{-11}_{33}(K)\,\gamma^\nu\,G^{-11}_{11}(K-P)+\gamma^\mu\,G^{-22}_{33}(K)\,\gamma^\nu\,G^{-22}_{11}(K-P)+\gamma^\mu\,G^{-33}_{33}(K)\,\gamma^\nu\,G^{-33}_{11}(K-P)\right.\nonumber\\
&+&\left. \gamma^\mu\,\Xi^{-13}_{31}(K)\,\gamma^\nu\,\Xi^{+31}_{31}(K-P)+\gamma^\mu\,\Xi^{-31}_{31}(K)\,\gamma^\nu\,\Xi^{+13}_{31}(K-P) \right.\nonumber\\
&+& \left. \gamma^\mu\,\Xi^{-13}_{13}(K)\,\gamma^\nu\,\Xi^{+31}_{13}(K-P)+\gamma^\mu\,\Xi^{-31}_{13}(K)\,\gamma^\nu\,\Xi^{+13}_{13}(K-P)\right.\nonumber\\
&+&\left. \gamma^\mu\,\Xi^{+13}_{31}(K)\,\gamma^\nu\,\Xi^{-31}_{31}(K-P)+\gamma^\mu\,\Xi^{+31}_{31}(K)\,\gamma^\nu\,\Xi^{-13}_{31}(K-P)\right.\nonumber\\
&+&\left. \gamma^\mu\,\Xi^{+13}_{13}(K)\,\gamma^\nu\,\Xi^{-31}_{13}(K-P)+\gamma^\mu\,\Xi^{+31}_{13}(K)\,\gamma^\nu\,\Xi^{-13}_{13}(K-P)\,
\right]\;,
\eea
\bea
\Pi^{\mu\nu}_{66}(P)&=& \Pi^{\mu\nu}_{77}(P)= \frac{g^2T}{8V}\sum_K {\rm Tr}_{s}\left[\right.\nonumber\\
&+&\left.\gamma^\mu\,G^{+11}_{22}(K)\,\gamma^\nu\,G^{+11}_{33}(K-P)+\gamma^\mu\,G^{+22}_{22}(K)\,\gamma^\nu\,G^{+22}_{33}(K-P)+\gamma^\mu\,G^{+33}_{22}(K)\,\gamma^\nu\,G^{+33}_{33}(K-P)\right.\nonumber\\
&+&\left.\gamma^\mu\,G^{+11}_{33}(K)\,\gamma^\nu\,G^{+11}_{22}(K-P)+\gamma^\mu\,G^{+22}_{33}(K)\,\gamma^\nu\,G^{+22}_{22}(K-P)+\gamma^\mu\,G^{+33}_{33}(K)\,\gamma^\nu\,G^{+33}_{22}(K-P)\right.\nonumber\\
&+&\left.\gamma^\mu\,G^{-11}_{22}(K)\,\gamma^\nu\,G^{-11}_{33}(K-P)+\gamma^\mu\,G^{-22}_{22}(K)\,\gamma^\nu\,G^{-22}_{33}(K-P)+\gamma^\mu\,G^{-33}_{22}(K)\,\gamma^\nu\,G^{-33}_{33}(K-P)\right.\nonumber\\
&+&\left.\gamma^\mu\,G^{-11}_{33}(K)\,\gamma^\nu\,G^{-11}_{22}(K-P)+\gamma^\mu\,G^{-22}_{33}(K)\,\gamma^\nu\,G^{-22}_{22}(K-P)+\gamma^\mu\,G^{-33}_{33}(K)\,\gamma^\nu\,G^{-33}_{22}(K-P)\right.\nonumber\\
&+&\left. \gamma^\mu\,\Xi^{-23}_{32}(K)\,\gamma^\nu\,\Xi^{+32}_{32}(K-P)+\gamma^\mu\,\Xi^{-32}_{32}(K)\,\gamma^\nu\,\Xi^{+23}_{32}(K-P) \right.\nonumber\\
&+& \left. \gamma^\mu\,\Xi^{-23}_{23}(K)\,\gamma^\nu\,\Xi^{+32}_{23}(K-P)+\gamma^\mu\,\Xi^{-32}_{23}(K)\,\gamma^\nu\,\Xi^{+23}_{23}(K-P)\right.\nonumber\\
&+&\left. \gamma^\mu\,\Xi^{+23}_{32}(K)\,\gamma^\nu\,\Xi^{-32}_{32}(K-P)+\gamma^\mu\,\Xi^{+32}_{32}(K)\,\gamma^\nu\,\Xi^{-23}_{32}(K-P)\right.\nonumber\\
&+&\left. \gamma^\mu\,\Xi^{+23}_{23}(K)\,\gamma^\nu\,\Xi^{-32}_{23}(K-P)+\gamma^\mu\,\Xi^{+32}_{23}(K)\,\gamma^\nu\,\Xi^{-23}_{23}(K-P)\,
\right]\;,
\eea
\bea
\Pi^{\mu\nu}_{88}(P)&=&\frac{g^2T}{24V}\sum_K {\rm Tr}_{s}\left[\right.\nonumber\\
&+&\left.\gamma^\mu\,G^{+11}_{11}(K)\,\gamma^\nu\,G^{+11}_{11}(K-P)+\gamma^\mu\,G^{+22}_{11}(K)\,\gamma^\nu\,G^{+22}_{11}(K-P)+\gamma^\mu\,G^{+33}_{11}(K)\,\gamma^\nu\,G^{+33}_{11}(K-P)\right.\nonumber\\
&+&\left.\gamma^\mu\,G^{+12}_{12}(K)\,\gamma^\nu\,G^{+21}_{21}(K-P)-2\gamma^\mu\,G^{+13}_{13}(K)\,\gamma^\nu\,G^{+31}_{31}(K-P)\right.\nonumber\\
&+&\left.\gamma^\mu\,G^{+11}_{22}(K)\,\gamma^\nu\,G^{+11}_{22}(K-P)+\gamma^\mu\,G^{+22}_{22}(K)\,\gamma^\nu\,G^{+22}_{22}(K-P)+\gamma^\mu\,G^{+33}_{22}(K)\,\gamma^\nu\,G^{+33}_{22}(K-P)\right.\nonumber\\
&+&\left.\gamma^\mu\,G^{+21}_{21}(K)\,\gamma^\nu\,G^{+12}_{12}(K-P)-2\gamma^\mu\,G^{+23}_{23}(K)\,\gamma^\nu\,G^{+32}_{32}(K-P)\right.\nonumber\\
&+&\left.4\,\gamma^\mu\,G^{+11}_{33}(K)\,\gamma^\nu\,G^{+11}_{33}(K-P)+4\,\gamma^\mu\,G^{+22}_{33}(K)\,\gamma^\nu\,G^{+22}_{33}(K-P)+4\,\gamma^\mu\,G^{+33}_{33}(K)\,\gamma^\nu\,G^{+33}_{33}(K-P)\right.\nonumber\\
&-&\left.2\gamma^\mu\,G^{+31}_{31}(K)\,\gamma^\nu\,G^{+13}_{13}(K-P)-2\gamma^\mu\,G^{+32}_{32}(K)\,\gamma^\nu\,G^{+23}_{23}(K-P)\right.\nonumber\\
&+&\left.\gamma^\mu\,G^{-11}_{11}(K)\,\gamma^\nu\,G^{-11}_{11}(K-P)+\gamma^\mu\,G^{-22}_{11}(K)\,\gamma^\nu\,G^{-22}_{11}(K-P)+\gamma^\mu\,G^{-33}_{11}(K)\,\gamma^\nu\,G^{-33}_{11}(K-P)\right.\nonumber\\
&+&\left.\gamma^\mu\,G^{-12}_{12}(K)\,\gamma^\nu\,G^{-21}_{21}(K-P)-2\gamma^\mu\,G^{-13}_{13}(K)\,\gamma^\nu\,G^{-31}_{31}(K-P)\right.\nonumber\\
&+&\left.\gamma^\mu\,G^{-11}_{22}(K)\,\gamma^\nu\,G^{-11}_{22}(K-P)+\gamma^\mu\,G^{-22}_{22}(K)\,\gamma^\nu\,G^{-22}_{22}(K-P)+\gamma^\mu\,G^{-33}_{22}(K)\,\gamma^\nu\,G^{-33}_{22}(K-P)\right.\nonumber\\
&+&\left.\gamma^\mu\,G^{-21}_{21}(K)\,\gamma^\nu\,G^{-12}_{12}(K-P)-2\gamma^\mu\,G^{-23}_{23}(K)\,\gamma^\nu\,G^{-32}_{32}(K-P)\right.\nonumber\\
&+&\left.4\,\gamma^\mu\,G^{-11}_{33}(K)\,\gamma^\nu\,G^{-11}_{33}(K-P)+4\,\gamma^\mu\,G^{-22}_{33}(K)\,\gamma^\nu\,G^{-22}_{33}(K-P)+4\,\gamma^\mu\,G^{-33}_{33}(K)\,\gamma^\nu\,G^{-33}_{33}(K-P)\right.\nonumber\\
&-&\left.2\gamma^\mu\,G^{-31}_{31}(K)\,\gamma^\nu\,G^{-13}_{13}(K-P)-2\gamma^\mu\,G^{-32}_{32}(K)\,\gamma^\nu\,G^{-23}_{23}(K-P)\right.\nonumber\\
&+& \left. \gamma^\mu\,\Xi^{-11}_{11}(K)\,\gamma^\nu\,\Xi^{+11}_{11}(K-P)+\gamma^\mu\,\Xi^{-22}_{22}(K)\,\gamma^\nu\,\Xi^{+22}_{22}(K-P)+4\,\gamma^\mu\,\Xi^{+33}_{33}(K)\,\gamma^\nu\,\Xi^{-33}_{33}(K-P)\right.\nonumber\\
&+&\left. \gamma^\mu\,\Xi^{-12}_{12}(K)\,\gamma^\nu\,\Xi^{+21}_{21}(K-P)+\gamma^\mu\,\Xi^{-21}_{12}(K)\,\gamma^\nu\,\Xi^{+12}_{21}(K-P) \right.\nonumber\\
&+& \left. \gamma^\mu\,\Xi^{-12}_{21}(K)\,\gamma^\nu\,\Xi^{+21}_{12}(K-P)+\gamma^\mu\,\Xi^{-21}_{21}(K)\,\gamma^\nu\,\Xi^{+12}_{12}(K-P)\right.\nonumber\\
&-&\left.2\, \gamma^\mu\,\Xi^{-13}_{13}(K)\,\gamma^\nu\,\Xi^{+31}_{31}(K-P)-2\,\gamma^\mu\,\Xi^{-31}_{13}(K)\,\gamma^\nu\,\Xi^{+13}_{31}(K-P) \right.\nonumber\\
&-& \left.2\, \gamma^\mu\,\Xi^{-13}_{31}(K)\,\gamma^\nu\,\Xi^{+31}_{13}(K-P)-2\,\gamma^\mu\,\Xi^{-31}_{31}(K)\,\gamma^\nu\,\Xi^{+13}_{13}(K-P)\right.\nonumber\\
&-&\left.2\, \gamma^\mu\,\Xi^{-23}_{23}(K)\,\gamma^\nu\,\Xi^{+32}_{32}(K-P)-2\,\gamma^\mu\,\Xi^{-32}_{23}(K)\,\gamma^\nu\,\Xi^{+23}_{32}(K-P) \right.\nonumber\\
&-& \left.2\, \gamma^\mu\,\Xi^{-23}_{32}(K)\,\gamma^\nu\,\Xi^{+32}_{23}(K-P)-2\,\gamma^\mu\,\Xi^{-32}_{32}(K)\,\gamma^\nu\,\Xi^{+23}_{23}(K-P)\right.\nonumber\\
&+& \left. \gamma^\mu\,\Xi^{+11}_{11}(K)\,\gamma^\nu\,\Xi^{-11}_{11}(K-P)+\gamma^\mu\,\Xi^{+22}_{22}(K)\,\gamma^\nu\,\Xi^{-22}_{22}(K-P)+4\,\gamma^\mu\,\Xi^{+33}_{33}(K)\,\gamma^\nu\,\Xi^{-33}_{33}(K-P)\right.\nonumber\\
&+&\left. \gamma^\mu\,\Xi^{+12}_{12}(K)\,\gamma^\nu\,\Xi^{-21}_{21}(K-P)+\gamma^\mu\,\Xi^{+21}_{12}(K)\,\gamma^\nu\,\Xi^{-12}_{21}(K-P)\right.\nonumber\\
&+&\left. \gamma^\mu\,\Xi^{+12}_{21}(K)\,\gamma^\nu\,\Xi^{-21}_{12}(K-P)+\gamma^\mu\,\Xi^{+21}_{21}(K)\,\gamma^\nu\,\Xi^{-12}_{12}(K-P)\right.\nonumber\\
&-&\left.2\, \gamma^\mu\,\Xi^{+13}_{13}(K)\,\gamma^\nu\,\Xi^{-31}_{31}(K-P)-2\,\gamma^\mu\,\Xi^{+31}_{13}(K)\,\gamma^\nu\,\Xi^{-13}_{31}(K-P)\right.\nonumber\\
&-&\left.2\, \gamma^\mu\,\Xi^{+13}_{31}(K)\,\gamma^\nu\,\Xi^{-31}_{13}(K-P)-2\,\gamma^\mu\,\Xi^{+31}_{31}(K)\,\gamma^\nu\,\Xi^{-13}_{13}(K-P)\right.\nonumber\\
&-&\left.2\, \gamma^\mu\,\Xi^{+23}_{23}(K)\,\gamma^\nu\,\Xi^{-32}_{32}(K-P)-2\,\gamma^\mu\,\Xi^{+32}_{23}(K)\,\gamma^\nu\,\Xi^{-23}_{32}(K-P)\right.\nonumber\\
&-&\left.2\, \gamma^\mu\,\Xi^{+23}_{32}(K)\,\gamma^\nu\,\Xi^{-32}_{23}(K-P)-2\,\gamma^\mu\,\Xi^{+32}_{32}(K)\,\gamma^\nu\,\Xi^{-23}_{23}(K-P)\,
\right]\;.
\eea
\end{subequations}
Furthermore, we see that, as before \cite{2frischke, 3frischke}, there are nonzero values for some of the the off-diagonal components
\begin{subequations}
\bea
\Pi^{\mu\nu}_{12}(P)&=& -\Pi^{\mu\nu}_{21}(P) = i\,\frac{g^2T}{8V}\sum_K {\rm Tr}_{s}\left[\right.\nonumber\\
&-&\left.\gamma^\mu\,G^{+11}_{11}(K)\,\gamma^\nu\,G^{+11}_{22}(K-P)-\gamma^\mu\,G^{+22}_{11}(K)\,\gamma^\nu\,G^{+22}_{22}(K-P)-\gamma^\mu\,G^{+33}_{11}(K)\,\gamma^\nu\,G^{+33}_{22}(K-P)\right.\nonumber\\
&+&\left.\gamma^\mu\,G^{+11}_{22}(K)\,\gamma^\nu\,G^{+11}_{11}(K-P)+\gamma^\mu\,G^{+22}_{22}(K)\,\gamma^\nu\,G^{+22}_{11}(K-P)+\gamma^\mu\,G^{+33}_{22}(K)\,\gamma^\nu\,G^{+33}_{11}(K-P)\right.\nonumber\\
&+&\left.\gamma^\mu\,G^{-11}_{11}(K)\,\gamma^\nu\,G^{-11}_{22}(K-P)+\gamma^\mu\,G^{-22}_{11}(K)\,\gamma^\nu\,G^{-22}_{22}(K-P)+\gamma^\mu\,G^{-33}_{11}(K)\,\gamma^\nu\,G^{-33}_{22}(K-P)\right.\nonumber\\
&-&\left.\gamma^\mu\,G^{-11}_{22}(K)\,\gamma^\nu\,G^{-11}_{11}(K-P)-\gamma^\mu\,G^{-22}_{22}(K)\,\gamma^\nu\,G^{-22}_{11}(K-P)-\gamma^\mu\,G^{-33}_{22}(K)\,\gamma^\nu\,G^{-33}_{11}(K-P)\right.\nonumber\\
&+&\left. \gamma^\mu\,\Xi^{-12}_{21}(K)\,\gamma^\nu\,\Xi^{+21}_{21}(K-P)+\gamma^\mu\,\Xi^{-21}_{21}(K)\,\gamma^\nu\,\Xi^{+12}_{21}(K-P) \right.\nonumber\\
&-& \left. \gamma^\mu\,\Xi^{-12}_{12}(K)\,\gamma^\nu\,\Xi^{+21}_{12}(K-P)-\gamma^\mu\,\Xi^{-21}_{12}(K)\,\gamma^\nu\,\Xi^{+12}_{12}(K-P)\right.\nonumber\\
&+&\left. \gamma^\mu\,\Xi^{+12}_{21}(K)\,\gamma^\nu\,\Xi^{-21}_{21}(K-P)+\gamma^\mu\,\Xi^{+21}_{21}(K)\,\gamma^\nu\,\Xi^{-12}_{21}(K-P)\right.\nonumber\\
&-&\left. \gamma^\mu\,\Xi^{+12}_{12}(K)\,\gamma^\nu\,\Xi^{-21}_{12}(K-P)-\gamma^\mu\,\Xi^{+21}_{12}(K)\,\gamma^\nu\,\Xi^{-12}_{12}(K-P)\,
\right]\;,
\eea
\bea
\Pi^{\mu\nu}_{38}(P)&=&\frac{g^2T}{8V\sqrt{3}}\sum_K {\rm Tr}_{s}\left[\right.\nonumber\\
&+&\left.\gamma^\mu\,G^{+11}_{11}(K)\,\gamma^\nu\,G^{+11}_{11}(K-P)+\gamma^\mu\,G^{+22}_{11}(K)\,\gamma^\nu\,G^{+22}_{11}(K-P)+\gamma^\mu\,G^{+33}_{11}(K)\,\gamma^\nu\,G^{+33}_{11}(K-P)\right.\nonumber\\
&+&\left.\gamma^\mu\,G^{+12}_{12}(K)\,\gamma^\nu\,G^{+21}_{21}(K-P)+2\,\gamma^\mu\,G^{+23}_{23}(K)\,\gamma^\nu\,G^{+32}_{32}(K-P)\right.\nonumber\\
&-&\left.\gamma^\mu\,G^{+11}_{22}(K)\,\gamma^\nu\,G^{+11}_{22}(K-P)-\gamma^\mu\,G^{+22}_{22}(K)\,\gamma^\nu\,G^{+22}_{22}(K-P)-\gamma^\mu\,G^{+33}_{22}(K)\,\gamma^\nu\,G^{+33}_{22}(K-P)\right.\nonumber\\
&-&\left.\gamma^\mu\,G^{+21}_{21}(K)\,\gamma^\nu\,G^{+12}_{12}(K-P)-2\,\gamma^\mu\,G^{+13}_{13}(K)\,\gamma^\nu\,G^{+31}_{31}(K-P)\right.\nonumber\\
&+&\left.\gamma^\mu\,G^{-11}_{11}(K)\,\gamma^\nu\,G^{-11}_{11}(K-P)+\gamma^\mu\,G^{-22}_{11}(K)\,\gamma^\nu\,G^{-22}_{11}(K-P)+\gamma^\mu\,G^{-33}_{11}(K)\,\gamma^\nu\,G^{-33}_{11}(K-P)\right.\nonumber\\
&+&\left.\gamma^\mu\,G^{-12}_{12}(K)\,\gamma^\nu\,G^{-21}_{21}(K-P)+2\,\gamma^\mu\,G^{-23}_{23}(K)\,\gamma^\nu\,G^{-32}_{32}(K-P)\right.\nonumber\\
&-&\left.\gamma^\mu\,G^{-11}_{22}(K)\,\gamma^\nu\,G^{-11}_{22}(K-P)-\gamma^\mu\,G^{-22}_{22}(K)\,\gamma^\nu\,G^{-22}_{22}(K-P)-\gamma^\mu\,G^{-33}_{22}(K)\,\gamma^\nu\,G^{-33}_{22}(K-P)\right.\nonumber\\
&-&\left.\gamma^\mu\,G^{-21}_{21}(K)\,\gamma^\nu\,G^{-12}_{12}(K-P)-2\,\gamma^\mu\,G^{-13}_{13}(K)\,\gamma^\nu\,G^{-31}_{31}(K-P)\right.\nonumber\\
&+&\left. \gamma^\mu\,\Xi^{-11}_{11}(K)\,\gamma^\nu\,\Xi^{+11}_{11}(K-P)+\gamma^\mu\,\Xi^{-12}_{12}(K)\,\gamma^\nu\,\Xi^{+21}_{21}(K-P)+\gamma^\mu\,\Xi^{-21}_{12}(K)\,\gamma^\nu\,\Xi^{+12}_{21}(K-P) \right.\nonumber\\
&-&\left.2\, \gamma^\mu\,\Xi^{-13}_{13}(K)\,\gamma^\nu\,\Xi^{+31}_{31}(K-P)-2\,\gamma^\mu\,\Xi^{-31}_{13}(K)\,\gamma^\nu\,\Xi^{+13}_{31}(K-P) \right.\nonumber\\
&-& \left. \gamma^\mu\,\Xi^{-22}_{22}(K)\,\gamma^\nu\,\Xi^{+22}_{22}(K-P)-\gamma^\mu\,\Xi^{-12}_{21}(K)\,\gamma^\nu\,\Xi^{+21}_{12}(K-P)-\gamma^\mu\,\Xi^{-21}_{21}(K)\,\gamma^\nu\,\Xi^{+12}_{12}(K-P)\right.\nonumber\\
&+& \left.2\, \gamma^\mu\,\Xi^{-23}_{23}(K)\,\gamma^\nu\,\Xi^{+32}_{32}(K-P)+2\,\gamma^\mu\,\Xi^{-32}_{23}(K)\,\gamma^\nu\,\Xi^{+23}_{32}(K-P)\right.\nonumber\\
&+&\left. \gamma^\mu\,\Xi^{+11}_{11}(K)\,\gamma^\nu\,\Xi^{-11}_{11}(K-P)+\gamma^\mu\,\Xi^{+12}_{12}(K)\,\gamma^\nu\,\Xi^{-21}_{21}(K-P)+\gamma^\mu\,\Xi^{+21}_{12}(K)\,\gamma^\nu\,\Xi^{-12}_{21}(K-P) \right.\nonumber\\
&-&\left.2\, \gamma^\mu\,\Xi^{+13}_{13}(K)\,\gamma^\nu\,\Xi^{-31}_{31}(K-P)-2\,\gamma^\mu\,\Xi^{+31}_{13}(K)\,\gamma^\nu\,\Xi^{-13}_{31}(K-P) \right.\nonumber\\
&-& \left. \gamma^\mu\,\Xi^{+22}_{22}(K)\,\gamma^\nu\,\Xi^{-22}_{22}(K-P)-\gamma^\mu\,\Xi^{+12}_{21}(K)\,\gamma^\nu\,\Xi^{-21}_{12}(K-P)-\gamma^\mu\,\Xi^{+21}_{21}(K)\,\gamma^\nu\,\Xi^{-12}_{12}(K-P)\right.\nonumber\\
&+& \left.2\, \gamma^\mu\,\Xi^{+23}_{23}(K)\,\gamma^\nu\,\Xi^{-32}_{32}(K-P)+2\,\gamma^\mu\,\Xi^{+32}_{23}(K)\,\gamma^\nu\,\Xi^{-23}_{32}(K-P)\,
\right]\;,
\eea
\bea
\Pi^{\mu\nu}_{83}(P)&=&\frac{g^2T}{8V\sqrt{3}}\sum_K {\rm Tr}_{s}\left[\right.\nonumber\\
&+&\left.\gamma^\mu\,G^{+11}_{11}(K)\,\gamma^\nu\,G^{+11}_{11}(K-P)+\gamma^\mu\,G^{+22}_{11}(K)\,\gamma^\nu\,G^{+22}_{11}(K-P)+\gamma^\mu\,G^{+33}_{11}(K)\,\gamma^\nu\,G^{+33}_{11}(K-P)\right.\nonumber\\
&-&\left.\gamma^\mu\,G^{+12}_{12}(K)\,\gamma^\nu\,G^{+21}_{21}(K-P)-2\,\gamma^\mu\,G^{+31}_{31}(K)\,\gamma^\nu\,G^{+13}_{13}(K-P)\right.\nonumber\\
&-&\left.\gamma^\mu\,G^{+11}_{22}(K)\,\gamma^\nu\,G^{+11}_{22}(K-P)-\gamma^\mu\,G^{+22}_{22}(K)\,\gamma^\nu\,G^{+22}_{22}(K-P)-\gamma^\mu\,G^{+33}_{22}(K)\,\gamma^\nu\,G^{+33}_{22}(K-P)\right.\nonumber\\
&+&\left.\gamma^\mu\,G^{+21}_{21}(K)\,\gamma^\nu\,G^{+12}_{12}(K-P)+2\,\gamma^\mu\,G^{+32}_{32}(K)\,\gamma^\nu\,G^{+23}_{23}(K-P)\right.\nonumber\\
&+&\left.\gamma^\mu\,G^{-11}_{11}(K)\,\gamma^\nu\,G^{-11}_{11}(K-P)+\gamma^\mu\,G^{-22}_{11}(K)\,\gamma^\nu\,G^{-22}_{11}(K-P)+\gamma^\mu\,G^{-33}_{11}(K)\,\gamma^\nu\,G^{-33}_{11}(K-P)\right.\nonumber\\
&-&\left.\gamma^\mu\,G^{-12}_{12}(K)\,\gamma^\nu\,G^{-21}_{21}(K-P)-2\,\gamma^\mu\,G^{-31}_{31}(K)\,\gamma^\nu\,G^{-13}_{13}(K-P)\right.\nonumber\\
&-&\left.\gamma^\mu\,G^{-11}_{22}(K)\,\gamma^\nu\,G^{-11}_{22}(K-P)-\gamma^\mu\,G^{-22}_{22}(K)\,\gamma^\nu\,G^{-22}_{22}(K-P)-\gamma^\mu\,G^{-33}_{22}(K)\,\gamma^\nu\,G^{-33}_{22}(K-P)\right.\nonumber\\
&+&\left.\gamma^\mu\,G^{-21}_{21}(K)\,\gamma^\nu\,G^{-12}_{12}(K-P)+2\,\gamma^\mu\,G^{-32}_{32}(K)\,\gamma^\nu\,G^{-23}_{23}(K-P)\right.\nonumber\\
&-&\left. \gamma^\mu\,\Xi^{-22}_{22}(K)\,\gamma^\nu\,\Xi^{+22}_{22}(K-P)-\gamma^\mu\,\Xi^{-12}_{12}(K)\,\gamma^\nu\,\Xi^{+21}_{21}(K-P)-\gamma^\mu\,\Xi^{-21}_{12}(K)\,\gamma^\nu\,\Xi^{+12}_{21}(K-P) \right.\nonumber\\
&-&\left.2\, \gamma^\mu\,\Xi^{-31}_{31}(K)\,\gamma^\nu\,\Xi^{+13}_{13}(K-P)-2\,\gamma^\mu\,\Xi^{-13}_{31}(K)\,\gamma^\nu\,\Xi^{+31}_{13}(K-P) \right.\nonumber\\
&+& \left. \gamma^\mu\,\Xi^{-11}_{11}(K)\,\gamma^\nu\,\Xi^{+11}_{11}(K-P)+\gamma^\mu\,\Xi^{-12}_{21}(K)\,\gamma^\nu\,\Xi^{+21}_{12}(K-P)+\gamma^\mu\,\Xi^{-21}_{21}(K)\,\gamma^\nu\,\Xi^{+12}_{12}(K-P)\right.\nonumber\\
&+& \left.2\, \gamma^\mu\,\Xi^{-32}_{32}(K)\,\gamma^\nu\,\Xi^{+23}_{23}(K-P)+2\,\gamma^\mu\,\Xi^{-23}_{32}(K)\,\gamma^\nu\,\Xi^{+32}_{23}(K-P)\right.\nonumber\\
&-&\left. \gamma^\mu\,\Xi^{+22}_{22}(K)\,\gamma^\nu\,\Xi^{-22}_{22}(K-P)-\gamma^\mu\,\Xi^{+12}_{12}(K)\,\gamma^\nu\,\Xi^{-21}_{21}(K-P)-\gamma^\mu\,\Xi^{+21}_{12}(K)\,\gamma^\nu\,\Xi^{-12}_{21}(K-P) \right.\nonumber\\
&-&\left.2\, \gamma^\mu\,\Xi^{+31}_{31}(K)\,\gamma^\nu\,\Xi^{-13}_{13}(K-P)-2\,\gamma^\mu\,\Xi^{+13}_{31}(K)\,\gamma^\nu\,\Xi^{-31}_{13}(K-P) \right.\nonumber\\
&+& \left. \gamma^\mu\,\Xi^{+11}_{11}(K)\,\gamma^\nu\,\Xi^{-11}_{11}(K-P)+\gamma^\mu\,\Xi^{+12}_{21}(K)\,\gamma^\nu\,\Xi^{-21}_{12}(K-P)+\gamma^\mu\,\Xi^{+21}_{21}(K)\,\gamma^\nu\,\Xi^{-12}_{12}(K-P)\right.\nonumber\\
&+& \left.2\, \gamma^\mu\,\Xi^{+32}_{32}(K)\,\gamma^\nu\,\Xi^{-23}_{23}(K-P)+2\,\gamma^\mu\,\Xi^{+23}_{32}(K)\,\gamma^\nu\,\Xi^{-32}_{23}(K-P)\,
\right]\;,
\eea
\bea
\Pi^{\mu\nu}_{45}(P)&=& -\Pi^{\mu\nu}_{54}(P) = i\,\frac{g^2T}{8V}\sum_K {\rm Tr}_{s}\left[\right.\nonumber\\
&-&\left.\gamma^\mu\,G^{+11}_{11}(K)\,\gamma^\nu\,G^{+11}_{33}(K-P)-\gamma^\mu\,G^{+22}_{11}(K)\,\gamma^\nu\,G^{+22}_{33}(K-P)-\gamma^\mu\,G^{+33}_{11}(K)\,\gamma^\nu\,G^{+33}_{33}(K-P)\right.\nonumber\\
&+&\left.\gamma^\mu\,G^{+11}_{33}(K)\,\gamma^\nu\,G^{+11}_{11}(K-P)+\gamma^\mu\,G^{+22}_{33}(K)\,\gamma^\nu\,G^{+22}_{11}(K-P)+\gamma^\mu\,G^{+33}_{33}(K)\,\gamma^\nu\,G^{+33}_{11}(K-P)\right.\nonumber\\
&+&\left.\gamma^\mu\,G^{-11}_{11}(K)\,\gamma^\nu\,G^{-11}_{33}(K-P)+\gamma^\mu\,G^{-22}_{11}(K)\,\gamma^\nu\,G^{-22}_{33}(K-P)+\gamma^\mu\,G^{-33}_{11}(K)\,\gamma^\nu\,G^{-33}_{33}(K-P)\right.\nonumber\\
&-&\left.\gamma^\mu\,G^{-11}_{33}(K)\,\gamma^\nu\,G^{-11}_{11}(K-P)-\gamma^\mu\,G^{-22}_{33}(K)\,\gamma^\nu\,G^{-22}_{11}(K-P)-\gamma^\mu\,G^{-33}_{33}(K)\,\gamma^\nu\,G^{-33}_{11}(K-P)\right.\nonumber\\
&+&\left. \gamma^\mu\,\Xi^{-13}_{31}(K)\,\gamma^\nu\,\Xi^{+31}_{31}(K-P)+\gamma^\mu\,\Xi^{-31}_{31}(K)\,\gamma^\nu\,\Xi^{+13}_{31}(K-P) \right.\nonumber\\
&-& \left. \gamma^\mu\,\Xi^{-13}_{13}(K)\,\gamma^\nu\,\Xi^{+31}_{13}(K-P)-\gamma^\mu\,\Xi^{-31}_{13}(K)\,\gamma^\nu\,\Xi^{+13}_{13}(K-P)\right.\nonumber\\
&-&\left. \gamma^\mu\,\Xi^{+13}_{31}(K)\,\gamma^\nu\,\Xi^{-31}_{31}(K-P)-\gamma^\mu\,\Xi^{+31}_{31}(K)\,\gamma^\nu\,\Xi^{-13}_{31}(K-P)\right.\nonumber\\
&+&\left. \gamma^\mu\,\Xi^{+13}_{13}(K)\,\gamma^\nu\,\Xi^{-31}_{13}(K-P)+\gamma^\mu\,\Xi^{+31}_{13}(K)\,\gamma^\nu\,\Xi^{-13}_{13}(K-P)\,
\right]\;,
\eea
\bea
\Pi^{\mu\nu}_{67}(P)&=& -\Pi^{\mu\nu}_{76}(P) = i\,\frac{g^2T}{8V}\sum_K {\rm Tr}_{s}\left[\right.\nonumber\\
&-&\left.\gamma^\mu\,G^{+11}_{22}(K)\,\gamma^\nu\,G^{+11}_{33}(K-P)-\gamma^\mu\,G^{+22}_{22}(K)\,\gamma^\nu\,G^{+22}_{33}(K-P)-\gamma^\mu\,G^{+33}_{22}(K)\,\gamma^\nu\,G^{+33}_{33}(K-P)\right.\nonumber\\
&+&\left.\gamma^\mu\,G^{+11}_{33}(K)\,\gamma^\nu\,G^{+11}_{22}(K-P)+\gamma^\mu\,G^{+22}_{33}(K)\,\gamma^\nu\,G^{+22}_{22}(K-P)+\gamma^\mu\,G^{+33}_{33}(K)\,\gamma^\nu\,G^{+33}_{22}(K-P)\right.\nonumber\\
&+&\left.\gamma^\mu\,G^{-11}_{22}(K)\,\gamma^\nu\,G^{-11}_{33}(K-P)+\gamma^\mu\,G^{-22}_{22}(K)\,\gamma^\nu\,G^{-22}_{33}(K-P)+\gamma^\mu\,G^{-33}_{22}(K)\,\gamma^\nu\,G^{-33}_{33}(K-P)\right.\nonumber\\
&-&\left.\gamma^\mu\,G^{-11}_{33}(K)\,\gamma^\nu\,G^{-11}_{22}(K-P)-\gamma^\mu\,G^{-22}_{33}(K)\,\gamma^\nu\,G^{-22}_{22}(K-P)-\gamma^\mu\,G^{-33}_{33}(K)\,\gamma^\nu\,G^{-33}_{22}(K-P)\right.\nonumber\\
&+&\left. \gamma^\mu\,\Xi^{-23}_{32}(K)\,\gamma^\nu\,\Xi^{+32}_{32}(K-P)+\gamma^\mu\,\Xi^{-32}_{32}(K)\,\gamma^\nu\,\Xi^{+23}_{32}(K-P) \right.\nonumber\\
&-&\left.\gamma^\mu\,\Xi^{-23}_{23}(K)\,\gamma^\nu\,\Xi^{+32}_{23}(K-P)-\gamma^\mu\,\Xi^{-32}_{23}(K)\,\gamma^\nu\,\Xi^{+23}_{23}(K-P)\right.\nonumber\\
&-&\left.\gamma^\mu\,\Xi^{+23}_{32}(K)\,\gamma^\nu\,\Xi^{-32}_{32}(K-P)-\gamma^\mu\,\Xi^{+32}_{32}(K)\,\gamma^\nu\,\Xi^{-23}_{32}(K-P)\right.\nonumber\\
&+&\left.\gamma^\mu\,\Xi^{+23}_{23}(K)\,\gamma^\nu\,\Xi^{-32}_{23}(K-P)+\gamma^\mu\,\Xi^{+32}_{23}(K)\,\gamma^\nu\,\Xi^{-23}_{23}(K-P)\,
\right]\;.
\eea
\end{subequations}
The other components are zero. As pointed out in Ref\,\cite{2frischke}, the off-diagonal components of the self-energy do not have any physical meanings. They simply indicate that the gluon propagator is not diagonal in the original basis of the adjoint colors. For instance, in the [12] subspace of adjoint colors, the resummed inverse gluon propagator, which is defined via
\bea
\Pi=\Delta^{-1} - \Delta_0^{-1}\;,
\eea
has the form
\bea\label{gprop}
\Delta^{-1}=\left(
\begin{array}{cc}
\Delta_0^{-1}+\Pi_{11} & i\,\hat{\Pi} \\
-i\,\hat{\Pi} & \Delta_0^{-1}+\Pi_{11}
\end{array}\right)\;,
\eea
where $\Delta_0^{-1}$ is the bare gluon propagator. The hermitian matrix of $\Delta^{-1}$ can be diagonalized by the unitary matrix
\bea
U\equiv\frac{1}{\sqrt{2}}\left(
\begin{array}{cc}
1 & -i \\
-i & 1
\end{array}\right)\;,
\eea
so that in the new basis of adjoint colors we have
\bea\label{diaggprop}
\Delta^{-1}=\left(
\begin{array}{cc}
\Delta_0^{-1}+\Pi_{11}+\hat{\Pi} & 0 \\
0 & \Delta_0^{-1}+\Pi_{11}-\hat{\Pi}
\end{array}\right)
=\left(\begin{array}{cc}
\Pi'_{11} & 0 \\
0 & \Pi'_{11}
\end{array}\right)\;.
\eea
The same argument applies for the [45] as well as the [67] subspaces. However, since $\Pi^{\mu\nu}_{38}\neq -\Pi^{\mu\nu}_{83}$, instead of \eq{gprop} we have
\bea
\Delta^{-1}=\left(
\begin{array}{cc}
\Delta_0^{-1}+\Pi_{33} & \hat{\Pi}_{38} \\
\hat{\Pi}_{83} & \Delta_0^{-1}+\Pi_{88}
\end{array}\right)\;,
\eea
which after diagonalizing, does not have the simple form of \eq{diaggprop} but
\bea\label{38matrix}
\Delta^{-1}=\frac{1}{2}\left(
\begin{array}{cc}
(x + w)+\sqrt{(x+w)^2-4(xw+yz)} & 0 \\
0 & (x + w)-\sqrt{(x+w)^2-4(xw+yz)}
\end{array}\right)
\equiv
\left(
\begin{array}{cc}
\Pi'_{33} & 0 \\
0 & \Pi'_{88}
\end{array}\right)\;,
\eea
where
\bea
x\equiv\Delta_0^{-1}+\Pi^{\mu\nu}_{33}\,\, , \,\, w\equiv\Delta_0^{-1}+\Pi^{\mu\nu}_{88}\,\, , \,\,
y\equiv\Pi^{\mu\nu}_{38}\,\, , \,\,z\equiv\Pi^{\mu\nu}_{83}\;\;.
\eea
As we see it is almost impossible to find analytically $\Pi'^{\mu\nu}_{33}$ and $\Pi'^{\mu\nu}_{88}$ from $\Pi_{33}^{\mu\nu}$ and $\Pi_{88}^{\mu\nu}$. There is another possibility. One, first, finds the values of $\Pi_{33}^{\mu\nu}$, $\Pi_{88}^{\mu\nu}$, $\Pi_{38}^{\mu\nu}$, and $\Pi_{83}^{\mu\nu}$ for certain values of momentum and energy, and afterwards, replaces them in \eq{38matrix} to find $\Pi'^{\mu\nu}_{33}$ and $\Pi'^{\mu\nu}_{88}$. Nevertheless, when we evaluate the self-energies in the CFL phase we recognize that $\Pi_{38}^{\mu\nu}$ and $\Pi_{83}^{\mu\nu}$ vanish, hence, no need to pursue the mentioned method. In that case the [38] subspace itself is diagonal, so that $\Pi_{33}^{\mu\nu}=\Pi'^{\mu\nu}_{33}$ and $\Pi_{88}^{\mu\nu}=\Pi'^{\mu\nu}_{88}$.

In the new bases, the self-energies are
\begin{subequations}\label{gaplessselfenergies}
\bea
\Pi'^{\mu\nu}_{11}(P)&=& \Pi'^{\mu\nu}_{22}(P)= \frac{g^2T}{4V}\sum_K {\rm Tr}_{s}\left[\right.\nonumber\\
&+&\left.\gamma^\mu\,G^{+11}_{22}(K)\,\gamma^\nu\,G^{+11}_{11}(K-P)+\gamma^\mu\,G^{+22}_{22}(K)\,\gamma^\nu\,G^{+22}_{11}(K-P)+\gamma^\mu\,G^{+33}_{22}(K)\,\gamma^\nu\,G^{+33}_{11}(K-P)\right.\nonumber\\
&+&\left.\gamma^\mu\,G^{-11}_{11}(K)\,\gamma^\nu\,G^{-11}_{22}(K-P)+\gamma^\mu\,G^{-22}_{11}(K)\,\gamma^\nu\,G^{-22}_{22}(K-P)+\gamma^\mu\,G^{-33}_{11}(K)\,\gamma^\nu\,G^{-33}_{22}(K-P)\right.\nonumber\\
&+&\left. \gamma^\mu\,\Xi^{-12}_{21}(K)\,\gamma^\nu\,\Xi^{+21}_{21}(K-P)+\gamma^\mu\,\Xi^{-21}_{21}(K)\,\gamma^\nu\,\Xi^{+12}_{21}(K-P) \right.\nonumber\\
&+&\left. \gamma^\mu\,\Xi^{+12}_{12}(K)\,\gamma^\nu\,\Xi^{-21}_{12}(K-P)+\gamma^\mu\,\Xi^{+21}_{12}(K)\,\gamma^\nu\,\Xi^{-12}_{12}(K-P)\,
\right]\;,
\eea
\bea
\Pi'^{\mu\nu}_{44}(P)&=& \Pi'^{\mu\nu}_{66}(P)= \frac{g^2T}{4V}\sum_K {\rm Tr}_{s}\left[\right.\nonumber\\
&+&\left.\gamma^\mu\,G^{+11}_{33}(K)\,\gamma^\nu\,G^{+11}_{11}(K-P)+\gamma^\mu\,G^{+22}_{33}(K)\,\gamma^\nu\,G^{+22}_{11}(K-P)+\gamma^\mu\,G^{+33}_{33}(K)\,\gamma^\nu\,G^{+33}_{11}(K-P)\right.\nonumber\\
&+&\left.\gamma^\mu\,G^{-11}_{11}(K)\,\gamma^\nu\,G^{-11}_{33}(K-P)+\gamma^\mu\,G^{-22}_{11}(K)\,\gamma^\nu\,G^{-22}_{33}(K-P)+\gamma^\mu\,G^{-33}_{11}(K)\,\gamma^\nu\,G^{-33}_{33}(K-P)\right.\nonumber\\
&+&\left. \gamma^\mu\,\Xi^{-13}_{31}(K)\,\gamma^\nu\,\Xi^{+31}_{31}(K-P)+\gamma^\mu\,\Xi^{-31}_{31}(K)\,\gamma^\nu\,\Xi^{+13}_{31}(K-P) \right.\nonumber\\
&+&\left. \gamma^\mu\,\Xi^{+13}_{13}(K)\,\gamma^\nu\,\Xi^{-31}_{13}(K-P)+\gamma^\mu\,\Xi^{+31}_{13}(K)\,\gamma^\nu\,\Xi^{-13}_{13}(K-P)\,
\right]\;,
\eea
\bea
\Pi'^{\mu\nu}_{55}(P)&=& \Pi'^{\mu\nu}_{77}(P)= \frac{g^2T}{4V}\sum_K {\rm Tr}_{s}\left[\right.\nonumber\\
&+&\left.\gamma^\mu\,G^{+11}_{33}(K)\,\gamma^\nu\,G^{+11}_{22}(K-P)+\gamma^\mu\,G^{+22}_{33}(K)\,\gamma^\nu\,G^{+22}_{22}(K-P)+\gamma^\mu\,G^{+33}_{33}(K)\,\gamma^\nu\,G^{+33}_{22}(K-P)\right.\nonumber\\
&+&\left.\gamma^\mu\,G^{-11}_{22}(K)\,\gamma^\nu\,G^{-11}_{33}(K-P)+\gamma^\mu\,G^{-22}_{22}(K)\,\gamma^\nu\,G^{-22}_{33}(K-P)+\gamma^\mu\,G^{-33}_{22}(K)\,\gamma^\nu\,G^{-33}_{33}(K-P)\right.\nonumber\\
&+&\left. \gamma^\mu\,\Xi^{-23}_{32}(K)\,\gamma^\nu\,\Xi^{+32}_{32}(K-P)+\gamma^\mu\,\Xi^{-32}_{32}(K)\,\gamma^\nu\,\Xi^{+23}_{32}(K-P) \right.\nonumber\\
&+&\left. \gamma^\mu\,\Xi^{+23}_{23}(K)\,\gamma^\nu\,\Xi^{-32}_{23}(K-P)+\gamma^\mu\,\Xi^{+32}_{23}(K)\,\gamma^\nu\,\Xi^{-23}_{23}(K-P)\,
\right]\;,
\eea
\end{subequations}
where we have used the property $\Pi^{\mu\nu}_{ij}(P)=\Pi^{\mu\nu}_{ij}(-P)$. Here we did not present the transformed version of $\Pi^{\mu\nu}_{33}(P)$ and $\Pi^{\mu\nu}_{88}(P)$; the reason is already explained.

To complete the calculations in the gCFL phase, one has to find the trace over the spinor space as well. This can be a subject for another article. For the moment, we are interested, in particular, in the self-energies of the CFL phase and we avoid the complications of the gCFL phase.

\section{Gluon Self-energies in CFL phase}\label{sec5}

Now we concentrate on the self-energies in the CFL phase. Since the gaps in the CFL phase are equal, Eq.\,(\ref{allgaps}), and also there are simple relations between the chemical potentials in this phase, Eq.\,(\ref{mus}), Eqs.\,(\ref{gaplessselfenergies}) find simpler forms. Making use of \eqq{cflprop}{cfloffprop}, Eqs.\,(\ref{gaplessselfenergies}) change to
\begin{subequations}\label{beforemixed}
\bea
\Pi'^{\mu\nu}_{11}(P)&=& \Pi'^{\mu\nu}_{22}(P)= \frac{g^2T}{4V}\sum_K {\rm Tr}_{s}\left[\right.\nonumber\\
&+&\left.\gamma^\mu\,G^{+22}_{11}(K)\,\gamma^\nu\,G^{+11}_{11}(K-P)+\gamma^\mu\,G^{+11}_{11}(K)\,\gamma^\nu\,G^{+22}_{11}(K-P)+\gamma^\mu\,G^{+33}_{11}(K)\,\gamma^\nu\,G^{+33}_{11}(K-P)\right.\nonumber\\
&+&\left.\gamma^\mu\,G^{-22}_{11}(K)\,\gamma^\nu\,G^{-11}_{11}(K-P)+\gamma^\mu\,G^{-11}_{11}(K)\,\gamma^\nu\,G^{-22}_{11}(K-P)+\gamma^\mu\,G^{-33}_{11}(K)\,\gamma^\nu\,G^{-33}_{11}(K-P)\right.\nonumber\\
&+&\left. \gamma^\mu\,\Xi^{-12}_{21}(K)\,\gamma^\nu\,\Xi^{+12}_{12}(K-P)+\gamma^\mu\,\Xi^{-12}_{12}(K)\,\gamma^\nu\,\Xi^{+12}_{21}(K-P) \right.\nonumber\\
&+&\left. \gamma^\mu\,\Xi^{+12}_{12}(K)\,\gamma^\nu\,\Xi^{-12}_{21}(K-P)+\gamma^\mu\,\Xi^{+12}_{21}(K)\,\gamma^\nu\,\Xi^{-12}_{12}(K-P)\,
\right]\;,
\eea
\bea
\Pi'^{\mu\nu}_{44}(P)&=& \Pi'^{\mu\nu}_{55}(P)= \Pi'^{\mu\nu}_{66}(P) = \Pi'^{\mu\nu}_{77}(P) = \frac{g^2T}{4V}\sum_K {\rm Tr}_{s}\left[\right.\nonumber\\
&+&\left.\gamma^\mu\,G^{+11}_{33}(K)\,\gamma^\nu\,G^{+11}_{11}(K-P)+\gamma^\mu\,G^{+11}_{33}(K)\,\gamma^\nu\,G^{+22}_{11}(K-P)+\gamma^\mu\,G^{+11}_{11}(K)\,\gamma^\nu\,G^{+33}_{11}(K-P)\right.\nonumber\\
&+&\left.\gamma^\mu\,G^{-11}_{11}(K)\,\gamma^\nu\,G^{-11}_{33}(K-P)+\gamma^\mu\,G^{-22}_{11}(K)\,\gamma^\nu\,G^{-11}_{33}(K-P)+\gamma^\mu\,G^{-33}_{11}(K)\,\gamma^\nu\,G^{-11}_{11}(K-P)\right.\nonumber\\
&+&\left. \gamma^\mu\,\Xi^{-13}_{31}(K)\,\gamma^\nu\,\Xi^{+12}_{12}(K-P)+\gamma^\mu\,\Xi^{-12}_{12}(K)\,\gamma^\nu\,\Xi^{+13}_{31}(K-P) \right.\nonumber\\
&+&\left. \gamma^\mu\,\Xi^{+12}_{12}(K)\,\gamma^\nu\,\Xi^{-31}_{13}(K-P)+\gamma^\mu\,\Xi^{+31}_{13}(K)\,\gamma^\nu\,\Xi^{-12}_{12}(K-P)\,
\right]\;,
\eea
\bea
\Pi^{\mu\nu}_{33}(P)&=&\frac{g^2T}{4V}\sum_K {\rm Tr}_{s}\left[\right.\nonumber\\
&+&\left.\gamma^\mu\,G^{+11}_{11}(K)\,\gamma^\nu\,G^{+11}_{11}(K-P)+\gamma^\mu\,G^{+22}_{11}(K)\,\gamma^\nu\,G^{+22}_{11}(K-P)+\gamma^\mu\,G^{+33}_{11}(K)\,\gamma^\nu\,G^{+33}_{11}(K-P)\right.\nonumber\\
&-&\left.\gamma^\mu\,G^{+12}_{12}(K)\,\gamma^\nu\,G^{+12}_{12}(K-P)\right.\nonumber\\
&+&\left.\gamma^\mu\,G^{-11}_{11}(K)\,\gamma^\nu\,G^{-11}_{11}(K-P)+\gamma^\mu\,G^{-22}_{11}(K)\,\gamma^\nu\,G^{-22}_{11}(K-P)+\gamma^\mu\,G^{-33}_{11}(K)\,\gamma^\nu\,G^{-33}_{11}(K-P)\right.\nonumber\\
&-&\left.\gamma^\mu\,G^{-12}_{12}(K)\,\gamma^\nu\,G^{-12}_{12}(K-P)\right.\nonumber\\
&-&\left. \gamma^\mu\,\Xi^{-12}_{12}(K)\,\gamma^\nu\,\Xi^{+12}_{12}(K-P)-\gamma^\mu\,\Xi^{-12}_{21}(K)\,\gamma^\nu\,\Xi^{+12}_{21}(K-P)+\gamma^\mu\,\Xi^{-11}_{11}(K)\,\gamma^\nu\,\Xi^{+11}_{11}(K-P) \right.\nonumber\\
&-&\left. \gamma^\mu\,\Xi^{+12}_{12}(K)\,\gamma^\nu\,\Xi^{-12}_{12}(K-P)-\gamma^\mu\,\Xi^{+12}_{21}(K)\,\gamma^\nu\,\Xi^{-12}_{21}(K-P)+\gamma^\mu\,\Xi^{+11}_{11}(K)\,\gamma^\nu\,\Xi^{-11}_{11}(K-P)\,
\right]\;,
\eea
\bea
\Pi^{\mu\nu}_{88}(P)&=&\frac{g^2T}{12V}\sum_K {\rm Tr}_{s}\left[\right.\nonumber\\
&+&\left.3\,\gamma^\mu\,G^{+11}_{11}(K)\,\gamma^\nu\,G^{+11}_{11}(K-P)+\gamma^\mu\,G^{+22}_{11}(K)\,\gamma^\nu\,G^{+22}_{11}(K-P)\right.\nonumber\\
&+&\left.\gamma^\mu\,G^{+33}_{11}(K)\,\gamma^\nu\,G^{+33}_{11}(K-P)
+ 4\,\gamma^\mu\,G^{+11}_{33}(K)\,\gamma^\nu\,G^{+11}_{33}(K-P)-3\,\gamma^\mu\,G^{+12}_{12}(K)\,\gamma^\nu\,G^{+12}_{12}(K-P)\right.\nonumber\\
&+&\left.3\,\gamma^\mu\,G^{-11}_{11}(K)\,\gamma^\nu\,G^{-11}_{11}(K-P)+\gamma^\mu\,G^{-22}_{11}(K)\,\gamma^\nu\,G^{-22}_{11}(K-P)\right.\nonumber\\
&+&\left.\gamma^\mu\,G^{-33}_{11}(K)\,\gamma^\nu\,G^{-33}_{11}(K-P)
+ 4\,\gamma^\mu\,G^{-11}_{33}(K)\,\gamma^\nu\,G^{-11}_{33}(K-P)-3\,\gamma^\mu\,G^{-12}_{12}(K)\,\gamma^\nu\,G^{-12}_{12}(K-P)\right.\nonumber\\
&-&\left.3\,\gamma^\mu\,\Xi^{-12}_{12}(K)\,\gamma^\nu\,\Xi^{+12}_{12}(K-P)+\gamma^\mu\,\Xi^{-12}_{21}(K)\,\gamma^\nu\,\Xi^{+12}_{21}(K-P) \right.\nonumber\\
&-&\left.2\,\gamma^\mu\,\Xi^{-31}_{13}(K)\,\gamma^\nu\,\Xi^{+13}_{31}(K-P)
- 2\, \gamma^\mu\,\Xi^{-13}_{31}(K)\,\gamma^\nu\,\Xi^{+31}_{13}(K-P)+ 3\, \gamma^\mu\,\Xi^{-11}_{11}(K)\,\gamma^\nu\,\Xi^{+11}_{11}(K-P)\right.\nonumber\\
&-&\left.3\,\gamma^\mu\,\Xi^{+12}_{12}(K)\,\gamma^\nu\,\Xi^{-12}_{12}(K-P)+\gamma^\mu\,\Xi^{+12}_{21}(K)\,\gamma^\nu\,\Xi^{-12}_{21}(K-P) \right.\nonumber\\
&-&\left.2\,\gamma^\mu\,\Xi^{+31}_{13}(K)\,\gamma^\nu\,\Xi^{-13}_{31}(K-P)
- 2\,\gamma^\mu\,\Xi^{+13}_{31}(K)\,\gamma^\nu\,\Xi^{-31}_{13}(K-P)+ 3\, \gamma^\mu\,\Xi^{+11}_{11}(K)\,\gamma^\nu\,\Xi^{-11}_{11}(K-P)\,
\right]\;.\nonumber\\
\eea
\end{subequations}
As expected, we see that in the CFL phase the off-diagonal components of the self-energies $\Pi^{\mu\nu}_{38}$ and $\Pi^{\mu\nu}_{83}$ vanish. Hence, we are again able to have all the self-energies in the diagonal bases.

To check the correctness of the our expressions we compare them with those given in Refs.\,\cite{3frischke, shovwij}, which are for the gluon self-energies in the CFL phase with $m_s=0$. By doing so, all the chemical potentials in Eqs.\,(\ref{mus}) become equal and this leads to
\begin{subequations}\label{identities}
\bea
G^{\pm 11}_{11}(K)=G^{\pm 22}_{22}(K)=G^{\pm 33}_{33}(K)\hspace{-.35cm}&&=\frac{1}{3}G^\pm_{\bf 1}(K)+\frac{2}{3}G^\pm_{\bf 8}(K)\\
G^{\pm 11}_{22}(K)=G^{\pm 22}_{11}(K)=G^{\pm 11}_{33}(K)=G^{\pm 33}_{11}(K)\hspace{-.35cm}&&=G^{\pm 22}_{33}(K)=G^{\pm 33}_{22}(K)=G^\pm_{\bf 8}(K)\\
G^{\pm 12}_{12}(K)=G^{\pm 21}_{21}(K)=G^{\pm 13}_{13}(K)=G^{\pm 31}_{31}(K)=G^{\pm 23}_{23}\hspace{-.4cm}&&(K)=G^{\pm 32}_{32}(K)=\frac{1}{3}G^\pm_{\bf 1}(K)-\frac{1}{3}G^\pm_{\bf 8}(K)
\eea
\end{subequations}
where $G^\pm_{\bf 1}(K)$ and $G^\pm_{\bf 8}(K)$ are the singlet and octet propagators, respectively. Moreover, we have
\begin{subequations}\label{identities}
\bea
\Xi^{\pm 11}_{11}(K)=\Xi^{\pm 22}_{22}(K)=\Xi^{\pm 33}_{33}(K)\hspace{-.35cm}&&=\frac{1}{3}\Xi^\pm_{\bf 1}(K)-\frac{2}{3}\Xi^\pm_{\bf 8}(K)\\
\Xi^{\pm 12}_{21}(K)=\Xi^{\pm 21}_{12}(K)=\Xi^{\pm 13}_{31}(K)=\Xi^{\pm 31}_{13}(K)\hspace{-.35cm}&&=\Xi^{\pm 23}_{32}(K)=\Xi^{\pm 32}_{23}(K)=-\Xi^\pm_{\bf 8}(K)\\
\Xi^{\pm 12}_{12}(K)=\Xi^{\pm 21}_{21}(K)=\Xi^{\pm 13}_{13}(K)=\Xi^{\pm 31}_{31}(K)=\Xi^{\pm 23}_{23}\hspace{-.4cm}&&(K)=\Xi^{\pm 32}_{32}(K)=\frac{1}{3}\Xi^\pm_{\bf 1}(K)+\frac{1}{3}\Xi^\pm_{\bf 8}(K)
\eea
\end{subequations}
Using these identities, we obtain the result of \cite{3frischke} that is
\bea
\Pi^{\mu\nu}_{ab}(P)=\delta_{ab}\,\Pi^{\mu\nu}(P)\,,
\eea
with
\bea
\Pi'^{\mu\nu}(P)&=& \frac{g^2T}{4V}\sum_K {\rm Tr}_{s}\left[\right.\nonumber\\
&+&\left.\gamma^\mu\,G^{+22}_{11}(K)\,\gamma^\nu\,G^{+11}_{11}(K-P)+\gamma^\mu\,G^{+11}_{11}(K)\,\gamma^\nu\,G^{+22}_{11}(K-P)+\gamma^\mu\,G^{+33}_{11}(K)\,\gamma^\nu\,G^{+33}_{11}(K-P)\right.\nonumber\\
&+&\left.\gamma^\mu\,G^{-22}_{11}(K)\,\gamma^\nu\,G^{-11}_{11}(K-P)+\gamma^\mu\,G^{-11}_{11}(K)\,\gamma^\nu\,G^{-22}_{11}(K-P)+\gamma^\mu\,G^{-33}_{11}(K)\,\gamma^\nu\,G^{-33}_{11}(K-P)\right.\nonumber\\
&+&\left. \gamma^\mu\,\Xi^{-12}_{21}(K)\,\gamma^\nu\,\Xi^{+12}_{12}(K-P)+\gamma^\mu\,\Xi^{-12}_{12}(K)\,\gamma^\nu\,\Xi^{+12}_{21}(K-P) \right.\nonumber\\
&+&\left. \gamma^\mu\,\Xi^{+12}_{12}(K)\,\gamma^\nu\,\Xi^{-12}_{21}(K-P)+\gamma^\mu\,\Xi^{+12}_{21}(K)\,\gamma^\nu\,\Xi^{-12}_{12}(K-P)\,
\right]\;.
\eea
For nonzero strange quark mass, the differences in the self-energies, Eqs.\,(\ref{beforemixed}), causes different screening masses for the associated gluons, see for instance Ref.\,\cite{gluonic3}.

To find the trace over the spinor space we have to introduce the mixed representations. For that, one has to Fourier transform and perform the Matsubara sum. In the next section, we present the form of the propagators after performing the sum over Matsubara frequencies.

\subsection{Mixed representation for quark propagators}

The mixed representation for the quark propagators is introduced via
\bea\label{mixedrep}
G^\pm(\tau, k)\equiv T\sum_{k_0} {\rm e}^{-k_0\tau}G^\pm(K) \hspace{.5cm} {\rm and} \hspace{.5cm} G^\pm(K)\equiv \int_0^{1/T} d\tau\, {\rm e}^{k_0\tau}G^\pm(\tau,k)\;.
\eea
Using these, one has
\bea\label{spinortrace}
T\sum_{k_0} {\rm Tr}_s\left[\,\gamma^\mu\, G^\pm(K_1)\,\gamma^\nu\, G^\pm(K_2)\,\right]=T\sum_{k_0}\int_0^{1/T} {\rm d}\tau_1 \,{\rm d}\tau_2 \,\,{\rm e}^{k_0\tau_1+(k_0-p_0)\tau_2}\,{\rm Tr}_s\left[\,\gamma^\mu\, G^\pm(\tau_1,k_1)\,\gamma^\nu\, G^\pm(\tau_2,k_2)\,\right]\;,
\eea
where the fermionic and the bosonic Matsubara frequencies are $k_0=-i(2n+1)\pi T$ and $p_0=-i2n\pi T$, respectively. To perform the Matsubara sum we use the identity
\bea
T\sum_m {\rm e}^{k_0\tau} = \sum_{m=-\infty}^{\infty}(-1)^m\delta\left(\tau-\frac{m}{T}\right)\;,
\eea
proved in Ref.\,\cite{lebellac}. Since $0\le \tau_1,\tau_2\le 1/T$, the identity is valid only for $m=1$, i.e. $\tau_2=1/T-\tau_1$. Furthermore, from Ref.\,\cite{2frischke} we have
\bea
G^\pm(-\tau,k)=\gamma_0\,G^\mp(\tau,k)\,\gamma_0\;.
\eea
Therefore, \eq{spinortrace} reads
\bea\label{donn}
T\sum_{k_0}{\rm Tr}_s\left[\,\gamma^\mu\, G^\pm(K_1)\,\gamma^\nu\, G^\pm(K_2)\,\right]=-\int_0^{1/T}{\rm d}\tau\,\,{\rm e}^{p_0\tau}\,{\rm Tr}_s\left[\,\gamma^\mu \,G^\pm(\tau_1,k_1)\,\gamma^\nu \,\gamma_0\, G^\mp(\tau_2,k_2)\,\gamma_0\,\right]\;.
\eea
We are now in the position to perform the Matsubara sum in terms of the contour integral in the complex $k_0$ plane for each propagator in \eq{cflprop}. After this, we have
\begin{subequations}
\bea
[G^\pm]^{11}_{11}=-\sum_{e=\pm}\hspace{-.3cm}&&\hspace{-.2cm}\left\{\frac{k_1\mp(\mu^{11}_{11}-ek)}{2k_1}\big[\,\theta(\tau)-N(+k_1)\,\big]\exp(-k_1\tau)\right.\nonumber\\
&+&\left. \frac{k_1\pm(\mu^{11}_{11}-ek)}{2k_1} \big[\,\theta(\tau)-N(-k_1)\,\big]\exp(+k_1\tau)\right\}\Lambda^\pm\gamma_0\;,
\eea
\bea
[G^\pm]^{11}_{22}=-\sum_{e=\pm}\hspace{-.3cm}&&\hspace{-.2cm}\left\{\frac{k_2\mp(\mu^{11}_{11}-ek)}{2k_2}\big[\,\theta(\tau)-N(+k_2)\,\big]\exp(-k_2\tau)\right.\nonumber\\
&+& \left.\frac{k_2\pm(\mu^{11}_{11}-ek)}{2k_2} \big[\,\theta(\tau)-N(-k_2)\,\big]\exp(+k_2\tau)\right\}\Lambda^\pm\gamma_0\;,
\eea
\bea
[G^\pm]^{11}_{33}=-\sum_{e=\pm}\hspace{-.3cm}&&\hspace{-.2cm}\left\{\frac{k^+_{3}+(\mu^{33}_{11}-ek)}{2(k^+_{3}-\frac{m_s^2}{2\mu})}\big[\,\theta(\tau)-N(\mp k^+_{3})\,\big]\exp(\pm k^+_{3}\tau)\right.\nonumber\\
&+&\left. \frac{k^-_{3}+(\mu^{33}_{11}-ek)}{2(k^-_{3}-\frac{m_s^2}{2\mu})} \big[\,\theta(\tau)-N(\mp k^-_{3})\,\big]\exp(\pm k^-_{3}\tau)\right\}\Lambda^\pm\gamma_0\;,
\eea
\bea
[G^\pm]^{33}_{11}=-\sum_{e=\pm}\hspace{-.3cm}&&\hspace{-.2cm}\left\{\frac{k^+_{3}-(\mu^{11}_{33}-ek)}{2(k^+_{3}-\frac{m_s^2}{2\mu})}\big[\,\theta(\tau)-N(\pm k^+_{3})\,\big]\exp(\mp k^+_{3}\tau)\right.\nonumber\\
&+&\left. \frac{k^-_{3}-(\mu^{11}_{33}-ek)}{2(k^-_{3}-\frac{m_s^2}{2\mu})} \big[\,\theta(\tau)-N(\pm k^-_{3})\,\big]\exp(\mp k^-_{3}\tau)\right\}\Lambda^\pm\gamma_0\;,
\eea
\end{subequations}
where $N(x)=\left(\exp(x/T)+1\right)^{-1}$. For the off-diagonal components of the propagator
\begin{subequations}
\bea
[\Xi^\pm]^{11}_{11}=\pm\sum_{e=\pm}\frac{\Lambda^\mp\gamma_5\sigma}{2k_1}\Big\{\big[\,\theta(\tau)-N(+k_1)\,\big]\exp(-k_1\tau)- \big[\,\theta(\tau)-N(-k_1)\,\big]\exp(+k_1\tau)\Big\}\;,
\eea
\bea
[\Xi^\pm]^{12}_{12}=\pm\sum_{e=\pm}\frac{\Lambda^\mp\gamma_5\varphi}{2k_1}\Big\{\big[\,\theta(\tau)-N(+k_1)\,\big]\exp(-k_1\tau)- \big[\,\theta(\tau)-N(-k_1)\,\big]\exp(+k_1\tau)\Big\}\;,
\eea
\bea
[\Xi^\pm]^{12}_{21}=\pm\sum_{e=\pm}\frac{\Lambda^\mp\gamma_5\phi}{2k_2}\Big\{\big[\,\theta(\tau)-N(+k_2)\,\big]\exp(-k_2\tau)- \big[\,\theta(\tau)-N(-k_2)\,\big]\exp(+k_2\tau)\Big\}\;,
\eea
\bea
[\Xi^\pm]^{31}_{13}=\pm\sum_{e=\pm}\frac{\Lambda^\mp\gamma_5\phi}{\,\,2(k^+_{3}-\frac{m_s^2}{2\mu})\,\,}\Big\{\big[\,\theta(\tau)-N(\mp k^-_{3})\,\big]\exp(\pm k^-_{3}\tau)- \big[\,\theta(\tau)-N(\mp k^+_{3})\,\big]\exp(\pm k^+_{3}\tau)\Big\}\;,
\eea
\bea
[\Xi^\pm]^{13}_{31}=\pm\sum_{e=\pm}\frac{\Lambda^\mp\gamma_5\phi}{\,\,2(k^+_{3}-\frac{m_s^2}{2\mu})\,\,}\Big\{\big[\,\theta(\tau)-N(\pm k^+_{3})\,\big]\exp(\mp k^+_{3}\tau)- \big[\,\theta(\tau)-N(\pm k^-_{3})\,\big]\exp(\mp k^-_{3}\tau)\Big\}\;.
\eea
\end{subequations}
In these expressions we have introduced
\begin{subequations}
\bea
k_1&=&\sqrt{(\mu^{11}_{11}-ek)^2 + 2\varphi^2+\sigma^2}\;,\\
k_2&=&\sqrt{(\mu^{11}_{11}-ek)^2 + \phi^2}\;,\\
k^+_{3}&=&\frac{m_s^2}{2\mu}+\sqrt{(\frac{m_s^2}{2\mu})^2+(\mu^{11}_{33}-ek)(\mu^{33}_{11}-ek)+ \phi^2}\;,\\
k^-_{3}&=&\frac{m_s^2}{2\mu}-\sqrt{(\frac{m_s^2}{2\mu})^2+(\mu^{11}_{33}-ek)(\mu^{33}_{11}-ek)+ \phi^2}\;.
\eea
\end{subequations}
Now, the procedure to find the spectral densities is very similar to the method indicated in Ref.\,\cite{3cf}. However, there is one difference. We see that the form of the propagators $[G^\pm]^{12}_{12}$ in Eq.\,(\ref{cflprop}) are different from the rest. This causes some anomalies which will be solved in Ref\,\cite{future}. For finding the gluon spectral densities, the only step needed to be taken is just to replace the above equations and do some numerical evaluations.

\section{Conclusions}

I enforced color and electric charge neutrality conditions on a system composed of three-flavor quark matter in the color superconducting phase. Under this condition I found the explicit form of the quark propagators and using them I derived the gluon self-energies in both the gCFL and the CFL phases. The results for the quark propagators can be used for many different purposes in the studies of thermodynamics of neutron stars. The form of the self-energies is in agreement with the previous results which indicate different screening masses for different gluons \cite{gluonic3}. These results will be, in particular, used to answer the question of the light plasmon condensation which might solve the chromomagnetic instabilities of the gCFL phase \cite{future}.

\section{Acknowledgments}

The author is indebted to M. Huang and D. H. Rischke for the useful discussions. He also thanks Frankfurt International Graduate School for Science and Institut f\"ur Theoretische Physik of Muenster university for financial supports.

\end{document}